\documentclass[12pt]{iopart}

\usepackage{graphicx}	
\usepackage{xcolor}
\usepackage{harvard}

\begin{document}
\title[Continuous Observations from the Ground]
{Continuous Solar Observations from the Ground -- Assessing Duty Cycle from GONG Observations}

\author{Kiran Jain, Sushant C. Tripathy, Frank Hill \& Alexei A. Pevtsov}

\address{National Solar Observatory, 3665 Discovery Dr., Boulder, CO 80303, USA}
\ead{kjain@nso.edu}

\begin{abstract}
Continuous observations play an important role in the studies of solar variability. 
While such observations can be achieved from space with 
 almost 100\% duty cycle, it is difficult to accomplish very high duty cycle from 
the ground. In this context, we assess the duty cycle that has been achieved from the
ground by analyzing the observations of a six station network of identical instruments, 
Global Oscillation Network Group (GONG). We provide a detailed analysis of the
duty cycle using GONG observations spanning over 18 years. We also  discuss duty 
cycle of individual sites and point out various factors that may impact individual site or 
network duty cycle. The mean duty cycle of the network
 is 93\%, however it reduces by about 5\% after all images pass through the stringent 
quality-control checks. The standard deviations in monthly and yearly duty cycle values
are found to be 1.9\% and 2.2\%, respectively. 
These results provide a baseline that can be used in the planning of future 
ground-based networks. 
\end{abstract}

\\
\submitto{Pub. Astron. Soc. Pacific}
\maketitle


\section{Introduction}
 \label{S-Introduction} 

Solar observations are now available for a few centuries, however most of the
records prior to the 20$^{\rm th}$ century 
are sparse.  The coverage has significantly increased in last 
few decades, primarily  due to the advancement in observing capabilities from both 
space and ground. While observations from space are almost continuous with small 
disruptions, the ground-based observations are adversely affected by the diurnal cycle.
There are several research areas in solar physics where uninterrupted observations for
a long time are essential in order to achieve reliable results. Helioseismology is
among one of these areas where solar oscillations below the surface are used to
study phenomena occurring in the  interior. These oscillations with a prominent
period of 5-minutes were discovered  in early 1960s \cite{leighton60,Leighton62} 
and the physical mechanism explaining these observations was suggested about a decade 
later \cite{Ulrich70,Leibacher71}.  It was soon realized that the continuous 
observations at high cadence for a longer time would be needed  to 
reduce side lobes in the temporal power spectra, but this was not 
possible from a single ground station due to the diurnal cycle. There were two 
possibilities for  continuous observations from ground: either from the South Pole 
but for a limited time during the 
Austral Summer \cite{Grec80} or  build a network of identical instruments at geographically 
separated locations around the Earth.  This led to the inception of several networks
for helioseismic observations; some 
are still operational even after more than 25 years and other ceased operations after a 
few years. Observations from these networks allowed the measurement of solar oscillations 
from low to high degrees where  degree is the number of waves around the solar circumference.

Some of the short-lived networks were the {\it International Research on the Interior of 
the Sun}  \cite[IRIS]{Fossat91} and the {\it Taiwanese Oscillations Network}   
\cite[TON]{Chou95}. Two still operating networks are the {\it Birmingham Solar Oscillation
Network} \cite[BISON]{Chaplin96} since 1975 and the {\it Global Oscillation Network Group}
 \cite[GONG]{Leibacher95} since 1995. While BiSON provides observations in 
 solar-disk-integrated  light best suitable for low spherical harmonic degree helioseismic 
 modes, which travel deep into the core, GONG observations resolve acoustic modes covering a 
 wider range from core to the near-surface layers. In addition, several observational 
 campaigns were held at the South Pole from the early 1980s to mid-1990s \cite{Harvey13,Fossat13},
 however continuous observations were obtained for a maximum period of about a 
 week. Occasional observation campaigns at the South Pole are still continuing with specific 
 science goals \cite{Finsterle04,Jefferies19} but continuous observations were acquired 
 only for a few days. All of these observations  played a crucial role in advancing the
 research in helioseismology and confirmed the solar variability below the surface
 as observed above the surface, first reported  by \citename{Woodard85} \citeyear{Woodard85}.  
 Furthermore, observations from the current or past space missions also complemented the 
 ground-based observations \cite{Palle15}. 

In this article, we assess the temporal coverage from the ground-based observatories
using observations of the GONG network.  Section~\ref{S-History} presents a brief history 
of GONG. Data used to compute duty cycle is described in Section~\ref{S-Data}. Assessment 
of the duty cycle of individual  sites and the network as a whole are discussed in
Sections~\ref{S-Sites} and \ref{S-Network}, respectively. We also provide statistical 
analysis of the simultaneous observations between different sites in \ref{S-Simul}. 
Section~\ref{S-Impact} demonstrates the scientific need of high duty cycle, especially
in helioseismic studies. Finally, the summary is presented in Section~\ref{S-Summary}.  
  
\section{A Brief History of GONG: From Helioseismology to Space Weather}
\label{S-History}


\begin{table}
\caption{\label{sites} Locations of GONG sites and operation start dates. }
\begin{indented}
\item[]\begin{tabular}{@{}lcccrc}     
\br                  
\multicolumn{2}{c}{Site} &  Latitude       &   Longitude     &   Elevation  \\
\cline{1-2}
  Name & Identifier        &                 &                 &    [meters]  \\
  \mr
Learmonth, Australia & LE       & S 022$^\circ$\,13$^\prime$\,06.6$^{\prime\prime}$& E
114$^\circ$\,06$^\prime$\,09.8$^{\prime\prime}$&  15\\
Udaipur, India       & UD       & N 024$^\circ$\,36$^\prime$\,53.8$^{\prime\prime}$& E
073$^\circ$\,40$^\prime$\,10.9$^{\prime\prime}$& 677\\
El Teide, Spain  & TD   & N 028$^\circ$\,18$^\prime$\,03.0$^{\prime\prime}$& W 
016$^\circ$\,30$^\prime$\,43.0$^{\prime\prime}$& 2425\\
Cerro Tololo, Chile  & CT         & S 030$^\circ$\,10$^\prime$\,04.2$^{\prime\prime}$& W
070$^\circ$\,48$^\prime$\,19.7$^{\prime\prime}$& 2190\\
Big Bear, USA    & BB           & N 034$^\circ$\,15$^\prime$\,37.2$^{\prime\prime}$& W
116$^\circ$\,55$^\prime$\,17.1$^{\prime\prime}$& 2063 \\
Mauna Loa, USA  &ML             & N 019$^\circ$\,32$^\prime$\,10.1$^{\prime\prime}$& W
155$^\circ$\,34$^\prime$\,33.3$^{\prime\prime}$& 3471 \\
  \mr
\end{tabular}
\end{indented}
\end{table}


\begin{figure}    
   \centerline{\includegraphics[width= 0.65\textwidth,clip=]{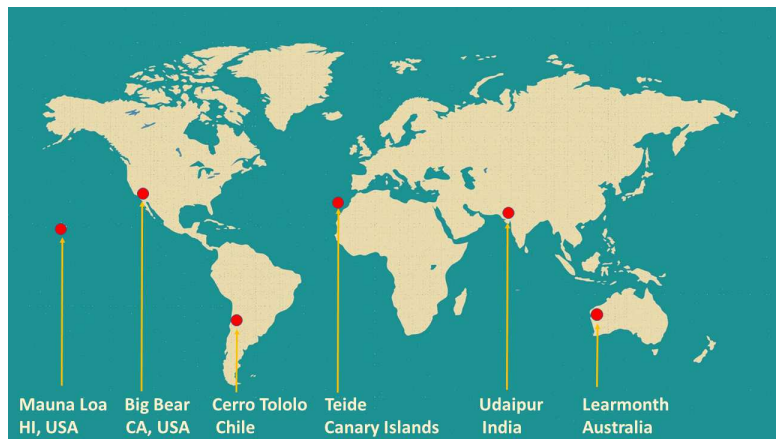}
              }
              \caption{Locations of GONG sites around the world.
   }\label{Sites}
   \end{figure}

One of the first tasks in establishing a network of observatories is to identify the 
number and locations of the sites suitable for substantially reducing the observed 
diurnal cycle. For this, a systematic study of potential sites for the GONG network 
started with estimating the performance of several hypothetical ground-based network 
composed of two to six existing observing sites distributed in both longitude and latitude.  
Results suggested that a network of six sites was needed to achieve $>$ 90\,\% annual mean 
duty cycle \cite{Hill85}. This initiated a coordinated site survey with identical 
instruments \cite{Fischer86} placed at 15 sites around the globe,  divided into six 
longitudinal bands to measure the cloud-cover percentage and transparency of the sky. Based 
on the detailed analysis of the collected data \cite{Hill94a} and the overall performance 
of each site \cite{Hill94b}, six sites were finally selected in early 1991. Each selected site
represents one of the six longitudinal bands that allows the network to make 24-hour-a-day 
observations of the Sun year round. These sites are shown in Figure~\ref{Sites} 
and their locations are given in Table~\ref{sites}.

The GONG instrument is based on measuring the line-of-sight Doppler shifts of the 
photospheric spectral line Ni {\sc i} 676.8\,nm.  The key element of this instrument is
a  Michelson interferometer, the details of which are given by \cite{harvey95}. Deployment 
of the GONG instruments started in January 1995 with the TD site and the full network was 
completed in October of the same year after UD, the last site, became operational.  Start 
dates of the operations at each site are  also given in Table~\ref{sites}. Identical
instruments and identical settings at all six sites operating in automated mode with 
minimal human intervention is the key to the network's success. The science-grade 
data of the  network are available since 7 May 1995. At the beginning, the
observations were carried out with low-resolution cameras that had rectangular pixels
(aspect ratio of 1.28:1) and was oriented in such a way that the longer dimension was 
aligned with the solar axis of rotation. The resulting images were 204 $\times$ 239 
pixels across with a pixel resolution of $\approx$10 $\times$ 8 arcseconds. 

After observing successfully for the initial three years using low-resolution cameras 
\cite{Harvey96}, the capabilities of the GONG network were expanded in 2001 by upgrading 
its cameras to 1024 $\times$ 1024 with square pixels \cite{Harvey98}. Until 2001, the GONG, 
commonly known as GONG classic,  data usage was primarily limited to global helioseismic
studies. The upgrade provided data that is suitable for local helioseismic studies, e.g. 
internal conditions below  small regions on the surface, hemispheric and latitudinal 
distributions of subsurface flows, mapping of the farside for the detection of active
regions, etc. The upgraded network is commonly referred to as GONG+. This also served as
the starting point to expand GONG's capabilities to space weather. In subsequent
years, the GONG observations were extended to one-minute cadence full-disk magnetograms in
2006 \cite{Petrie08,Hill08} and 2048 $\times$ 2048 pixels images of the H$\alpha$ spectral
line at a wavelength of 656.28 nm in mid-2010 \cite{Harvey11}.  Note that all H$\alpha$
observations are taken at a cadence of one minute at each site but with the acquisition time
offset between adjacent sites so that a new H$\alpha$ image is, in principle, available 
every 20 seconds from the network. Various GONG data products are available at
https://nso.edu/data/nisp-data/, and their applications in space-weather related studies 
and predictions are described by \cite{Hill18}.

\section{Data}
\label{S-Data}


\begin{table}
\caption{\label{nsite} Details of the  network data used in this study. }
\begin{indented}
\item[]\begin{tabular}{@{}lcrcc}     
  \br
Data Type        & Count & Duty Cycle & Details \\
\mr
Period covered                           &18 years                 &  & \\
                                         & (2002\,--\,2019)             & & \\
Total Number of days                     & 6574                    & & \\
Total Minutes covered                    & 9.47 $\times$ 10$^{6}$  & &  \\
Total Observations from all sites        & 1.35 $\times$ 10$^{7}$  & &  \\
\\
Total Minutes with observations         & 8.36  $\times$ 10$^{6}$  & 88.33\,\%  &\\
Total Minutes with no observations      & 1.11  $\times$ 10$^{6}$  & 11.67\,\%  &\\
\\
Total minutes with independent observations  & 4.12 $\times$ 10$^{6}$  &  43.49\,\%  & Table 3\\
Total minutes with simultaneous observations & 4.24 $\times$ 10$^{6}$  &  44.84\,\%  & Table 4\\
  \mr
\end{tabular}
\end{indented}
\end{table}

We compute the duty cycle of individual sites or the network from {\it fully calibrated
images} that have successfully passed through stringent quality checks as described in the
Automated Image Rejection \cite[AIR]{Clark04}, and consist of, for example, active tracking,
image stability,  signal/noise level in individual images \cite{Toussaint95,Pintar98},
temporal variations between images, and the correction for orientation for the best
position-angle estimation \cite{Toner01}.  As mentioned in Section~\ref{S-History},
 GONG has many different observables out of which we choose observations of Doppler velocities to
 determine the duty cycle. While the magnetograms and the Dopplergrams are part of the 
 same observing sequence, the data products go through different  processing pipelines
 and thus have different image rejection criteria as part of the quality control. 
 H$\alpha$ observations have been taken since 2010  using a different instrument setup and
 may provide  a slightly different duty cycle. 

 The duty cycle is defined as the fraction of 24-hour day for which the observations are
 available. In addition, the network duty cycle is determined by combining all six site  
 images for each minute in the same  24-hour period. There are several factors that
 significantly  influence the observations at individual sites and, finally, the network 
 duty cycle. These factors include, but are not limited to, severe weather conditions,
 instrument problems, downtime due to preventive maintenance and upgrades, etc. Although 
 the network has been operational for more than 25 years, we consider only a period of 18
 years from January 2002 to December 2019 for assessing the duty cycle, i.e. the period 
 after the completion of camera  upgrade. This period allows us to analyze consistent data
 set without any forced major disruptions.  A total number of 6574 days, i.e.
 9.47$\times$10$^6$ minutes, are covered  in this study. Since there are overlapping
 observations between different sites and  GONG provides observations at a cadence of one
 minute, the analysis has been carried out using more than 13 million images collected at
 GONG sites over the period of 18 years. Tables~\ref{nsite} summarizes the details of the
 network input data and Table~\ref{1site_contri} provides the  details of each site's data.
 In the next section, we  describe the performance  of individual sites starting from the
 East. Results for the full network are discussed  in Section~\ref{S-Network}.

\begin{table}
\caption{\label{1site_contri} Details of individual site data used in this study. }
\begin{indented}
\item[]\begin{tabular}{@{}ccccc}     
  \br                  
Sites & Contribution to total & \multicolumn{3}{c} {Contribution to network duty cycle} \\
\cline{3-5}
      & observations          & Total    & Independent & Overlapping \\
\mr
LE      & 18.6\,\%              &  26.6\,\%  & 12.9\,\%      & 13.7\,\%       \\
UD      & 11.2\,\%              &  15.9\,\%  & 5.1\,\%       & 10.8\,\%    \\
TD      & 17.8\,\%              &  25.7\,\%  & 8.7\,\%       & 17.0\,\%   \\
CT      & 21.1\,\%              &  30.1\,\%  & 8.0\,\%       & 22.1\,\%    \\
BB      & 16.5\,\%              &  23.6\,\%  & 4.7\,\%       & 18.9\,\%    \\
ML      & 14.8\,\%              &  21.2\,\%  & 4.1\,\%       & 17.1\,\%     \\
  \mr
\end{tabular}
\end{indented}
\end{table}
\section{Individual Site Duty Cycle}
\label{S-Sites}

\subsection{Learmonth, Australia (LE)}

\begin{figure}    
   \centerline{\includegraphics[width=0.6\textwidth,clip=]{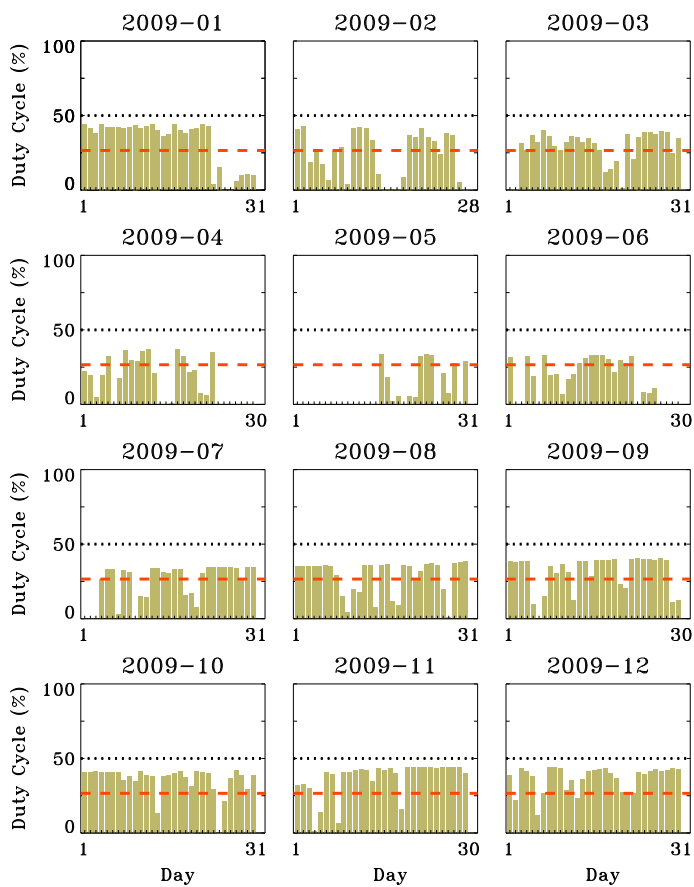}
              }
              \caption{Daily duty cycle for Learmonth (LE) site
for each month in the year 2009.  The dashed (red) line represents the mean
value over the entire period (see Table~\ref{site_summary}) whereas the dotted (black)
line shows a duty cycle value of 50\,\%.  
   }\label{daily_LE}
   \end{figure}

The Learmonth Solar Observatory (LSO) situated on the western shore of the Exmouth Gulf 
on the North West Cape of Australia houses one of  GONG's instruments.  The observatory 
is co-administered by the Bureau of Meteorology, Australian Government, and the US
Air Force. This GONG site became operational in April 1995. We display the variation of daily 
duty cycle in Figure~\ref{daily_LE}, as an example, for the year 2009. Here different 
panels represent different calendar months. In addition to the seasonal variation in 
duty cycle, there are some random gaps, which signify no 
observations for entire days. Longer gaps (e.g. from the last week of April to mid-May 
in Figure~\ref{daily_LE}) generally arise due to instrument downtime or preventive 
maintenance periods while shorter gaps correspond to severe weather that may last from one
to a few days. The maximum daily observations were obtained for 11.47 hours which gives 
a duty cycle of 47.8\,\%. In total, LE provided observations for 5621 days out of 6574 days 
(i.e. 85.5\,\% days). Various milestones  achieved by individual sites are given in 
Table~\ref{site_summary}.  

\begin{figure}    
   \centerline{\includegraphics[width=0.6\textwidth,clip=]{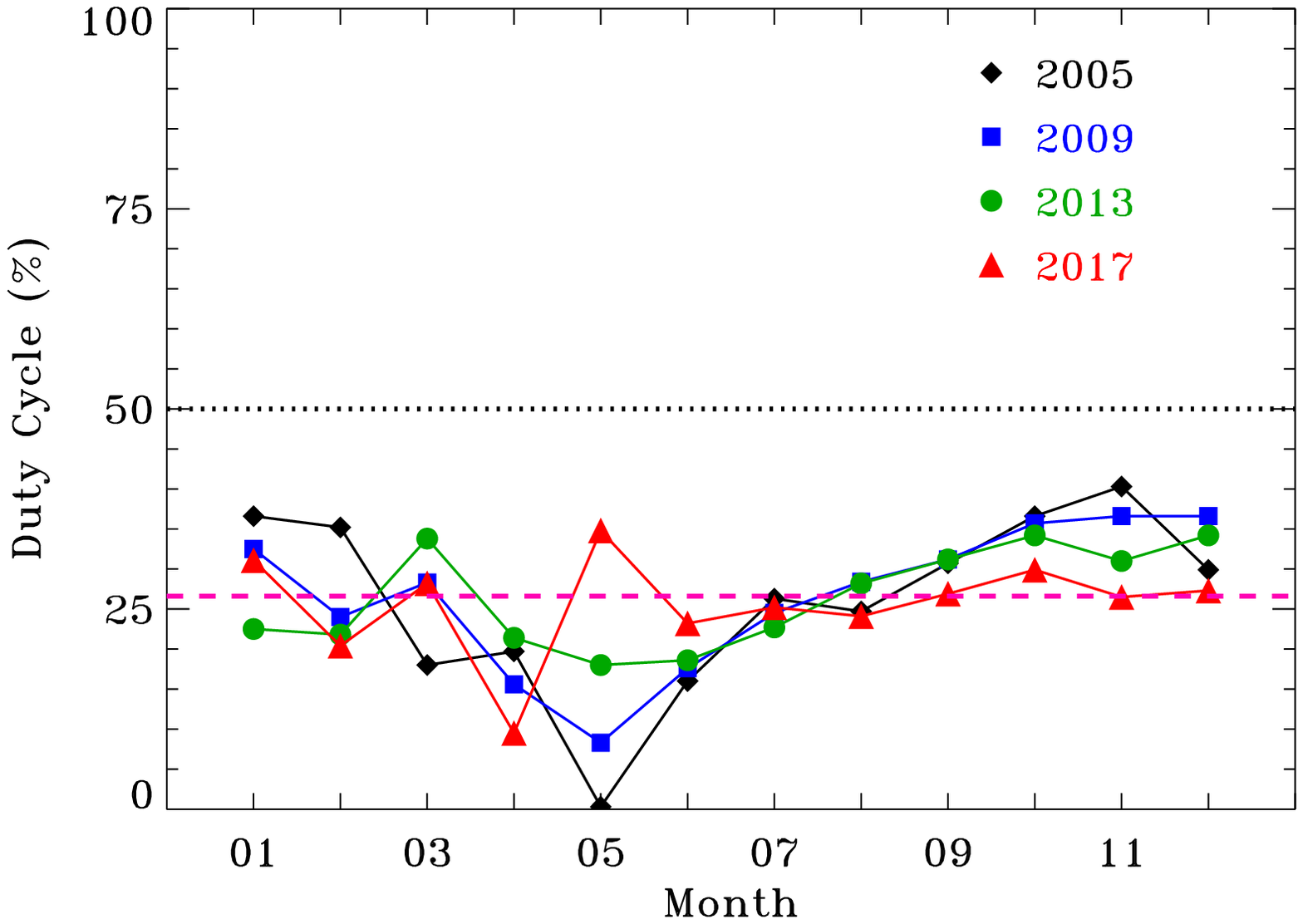}
              }
              \caption{Monthly duty cycle for Learmonth (LE) site for
selected years. Yearly means for  2005, 2009, 2013, and 2017
are 26\,\%, 27\,\%, 27\,\%, and 26\,\%, respectively.  The dashed (pink) line represents the mean
value over the entire period (see Table~\ref{site_summary}) whereas the dotted (black) line shows
a duty cycle value of 50\,\%. 
   }\label{monthly_LE}
   \end{figure}
\begin{table}
\caption{\label{site_summary}Various performance figures achieved by individual sites. }
\begin{indented}
\item[]\begin{tabular}{@{}llcccccccc}     
  \br                  
Criterion                 &Frequency & LE   & UD   & TD   & CT   & BB   & ML   \\
\mr
Total coverage [Days]    & All     & 5621   & 3949   & 5280   & 5516   & 5532   & 5536  \\
\\
Mean Duty Cycle [\%]          & All     & 26.6 & 16.0 & 25.4 & 30.2 & 23.6 & 21.2 \\
\\
Median Duty Cycle [\%]        & Monthly & 28.2  &18.4 & 26.5 & 31.1 & 25.5 &  21.4   \\
                         & Yearly  & 26.5  & 16.2 & 25.7 & 31.1 & 23.8 & 21.1 \\ 
                         \\
Maximum Duty Cycle [\%]      & Daily   & 47.8 & 46.7 & 49.4 & 50.3 & 48.5 & 47.7  \\
                         & Monthly & 34.2 & 27.0 & 39.4 & 43.6 & 36.3 & 26.1  \\
                         & Yearly  & 31.1 & 21.4 & 30.2 & 33.1 & 25.6 & 29.5  \\
\\
Minimum Duty Cycle [\%]       & Daily   & 0.0  & 0.0  & 0.0  & 0.0  & 0.0  & 0.0  \\
                         & Monthly & 17.8 & 4.0  & 16.1 & 18.0 & 12.6 & 18.1 \\
                         & Yearly  & 23.1 & 9.9  & 20.5 & 26.4 & 19.8 & 14.1  \\
\\
Standard Deviation [\%]      & Monthly & 6.0  & 9.7  & 8.3  & 9.7  & 7.3  & 2.6  \\
                         & Yearly  & 2.0  & 3.0  & 2.6  & 2.2  & 1.4  & 3.3  \\
  \mr
\end{tabular}
\end{indented}
\end{table}
\begin{figure}    
   \centerline{\includegraphics[width=0.75\textwidth,clip=]{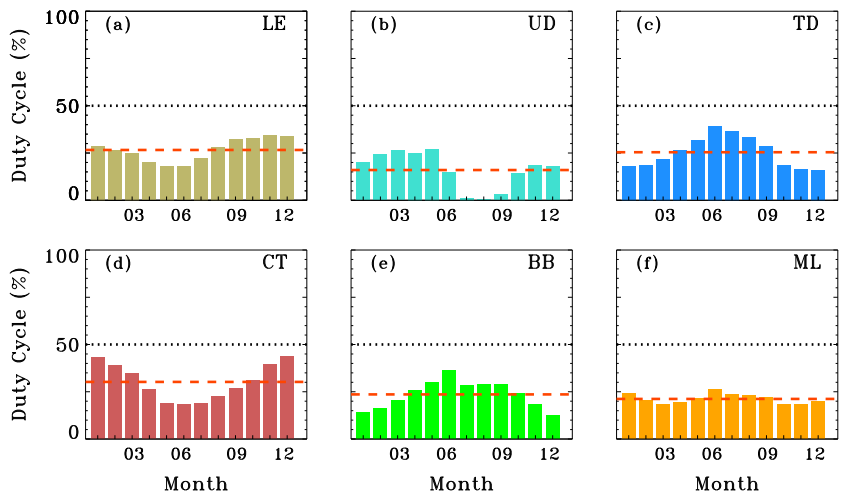}
              }
              \caption{Monthly variation in duty cycle averaged over the entire 
period for all six GONG sites. Horizontal dotted (black)  and  dashed (red) lines
represent 50\,\% and the mean value for individual sites over the entire period 
(see Table~\ref{site_summary}), respectively.
            }\label{site_monthly}
\end{figure}

\begin{figure}    
   \centerline{\includegraphics[width=0.75\textwidth,clip=]{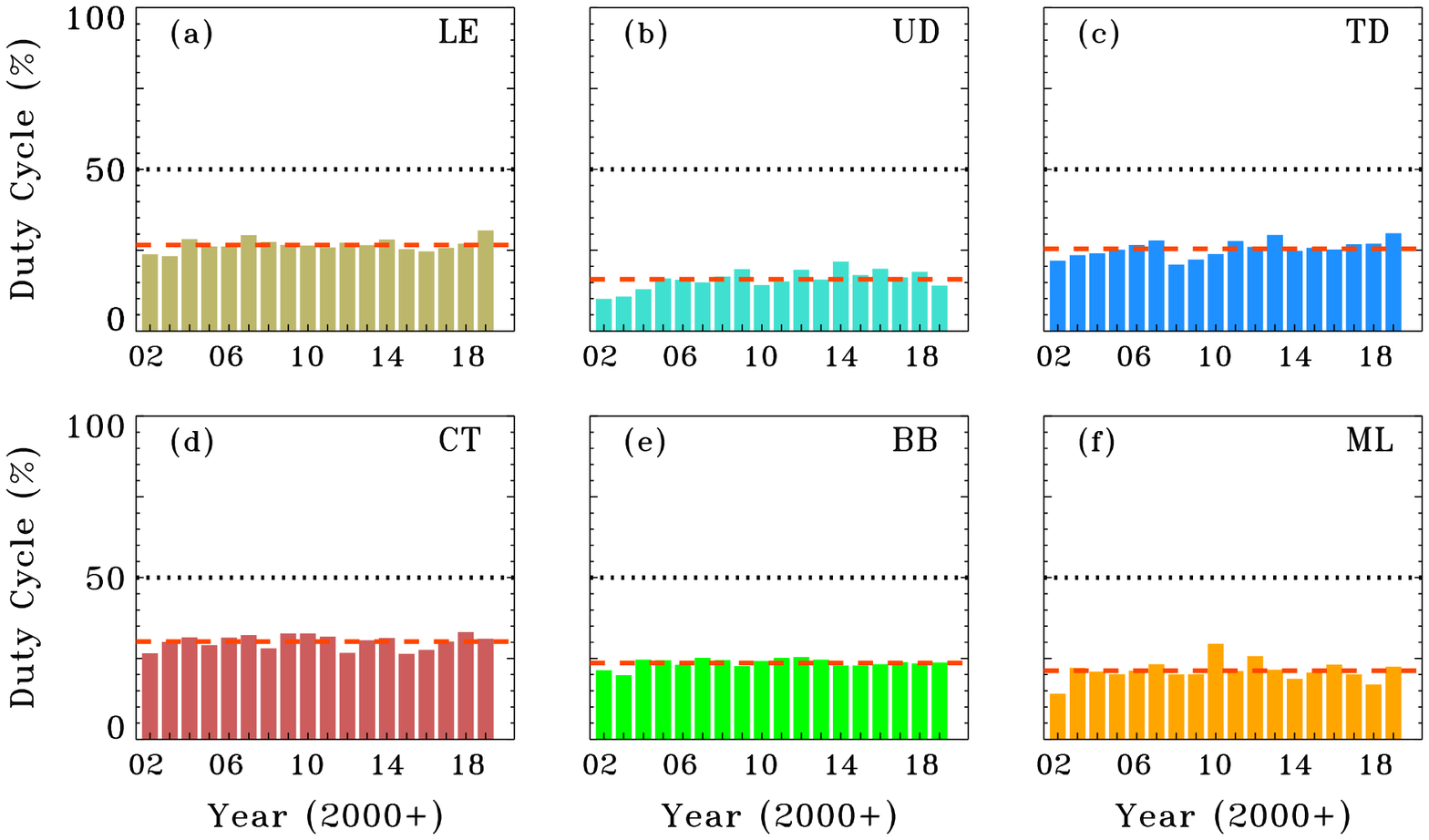}
              }
              \caption{Yearly variation in duty cycle for all six GONG sites.  
Horizontal dotted (black)  and  dashed (red) lines represent 50\,\% and 
the mean value for individual sites over the entire period 
(see Table~\ref{site_summary}), respectively.
            }\label{site_yearly}
\end{figure}

To examine the trends in LE observations, we display in Figure~\ref{monthly_LE}
the monthly variation in duty cycle for four selected years. A systematic seasonal
variation is seen in all four years with a dip in the duty cycle in April\,--\,May, 
except for 2017, followed by an increasing trend. The higher duty cycle is achieved 
in later months every year with the beginning of Summer in the southern hemisphere 
due to the availability of sunshine over a larger fraction of the day. The overall 
trends in monthly and yearly variations 
averaged over the entire period are presented in Figures~\ref{site_monthly} and 
\ref{site_yearly},  respectively.  We have included other GONG sites in the same figure 
for comparison. As mentioned above, the trend in the mean monthly values display strong 
seasonal variation with a standard deviation of 6.0\,\%. When we consider yearly variation 
(see Figure~\ref{site_yearly}), LE's contribution to the GONG network  has been stable 
throughout the entire period  and, as a result, the standard deviation reduces to about 2.0\,\%.

From Table~\ref{1site_contri}, we find that the observations taken at LE contribute about 
18.6\,\% to the total observations and 26.6\,\% to the network duty cycle.  As mentioned 
earlier, all sites provide independent as well as overlapping observations and LE's 
contribution to the network duty cycle is almost equally split between these two categories. 
It is further noted that LE's independent contribution is  the largest among all sites.

\subsection{Udaipur, India (UD)}

The Udaipur Solar Observatory (USO) is situated in a semi-arid highland region of Western 
India and houses the GONG's instrument in the Asian band. USO is administered by the Physical 
Research Laboratory (Ahmedabad), a unit of Department of Space, Government of India. The 
Udaipur site experiences strong weather pattern with an extended period of monsoon lasting 
for two\,--\,three months every year and the observations are stopped during this period. 
As a result, UD's contribution to the total observations is the lowest at  11.2\,\%.
As presented in Table~\ref{1site_contri}, UD  provides only 5.1\,\%  independent contribution 
to the network duty cycle while the total contribution is 15.9\,\%. Two-thirds of UD 
observations have overlap with other sites and the details are given in 
Section~\ref{S-Network}. 

\begin{figure}    
   \centerline{\includegraphics[width=0.6\textwidth,clip=]{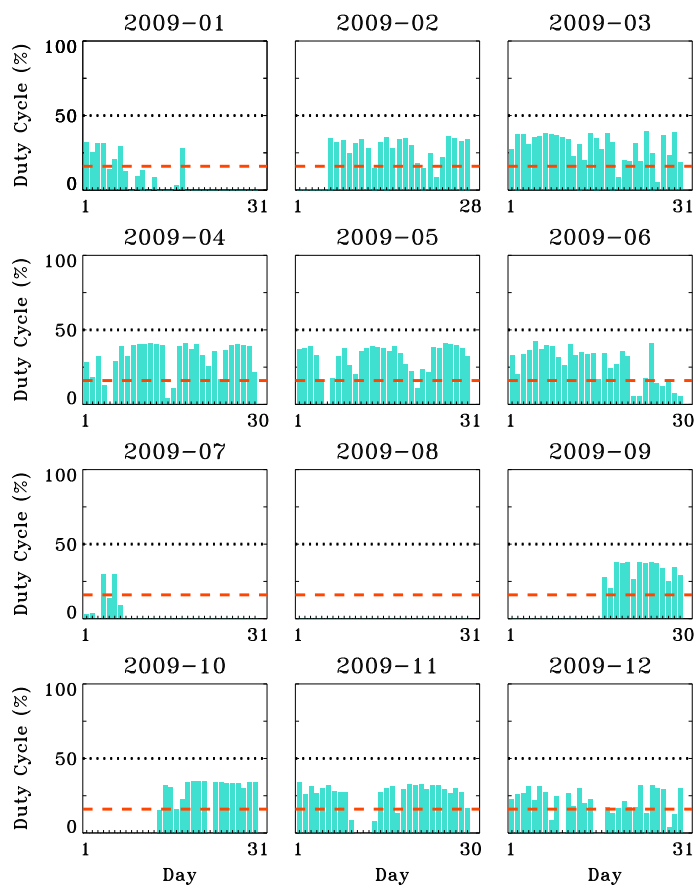}
              }
              \caption{Daily duty cycle for Udaipur (UD) site
for each month in the year 2009. The dashed (red) line represents the mean
value over the entire period (see Table~\ref{site_summary}) whereas the dotted (black)
line shows a duty cycle value of 50\,\%.  
   }\label{daily_UD}
   \end{figure}

\begin{figure}    
   \centerline{\includegraphics[width=0.6\textwidth,clip=]{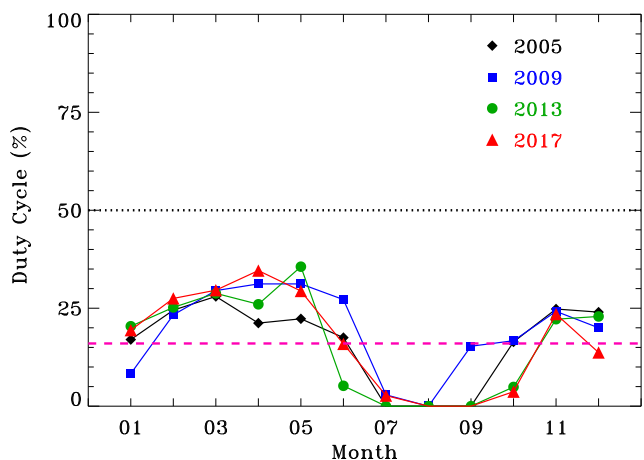}
              }
              \caption{Monthly duty cycle for Udaipur (UD) site for
selected years.  Yearly means for 2005, 2009, 2013 and 2017 are 16\,\%, 19\,\%, 16\,\%, 
and 17\,\%, respectively.  The dashed (pink) line represents the mean
value over the entire period (see Table~\ref{site_summary}) whereas the dotted (black)
line shows a duty cycle value of 50\,\%.  
   }\label{monthly_UD}
   \end{figure}

To evaluate the  daily variation throughout the year, we show daily duty cycle for the
year 2009 in  Figure~\ref{daily_UD} and the monthly duty cycle for four selected years in  
Figure~\ref{monthly_UD}. Compared to LE, Udaipur observations have many longer gaps in 2009.  
The UD site was down from early July to mid-September due to the monsoon and the other two 
gaps, in late January and early October, are due to system failures and occurrence of preventive 
maintenance. The downtime due to the monsoon is seen in all monthly average plots presented 
in Figure~\ref{monthly_UD}. The duty cycle again decreases in December\,--\,January due to 
hazy conditions and less sunshine hours available during Winter months. The maximum observations 
are normally obtained from February to May. During Spring/Summer time, the Udaipur site 
observed for more than 11 hours and the maximum daily duty cycle during 
2002\,--\,2019 reached 46.7\,\%.  

The average monthly and yearly duty-cycle trends, as shown in Figures~\ref{site_monthly}b 
and \ref{site_yearly}b, clearly demonstrate the effect of the monsoon on the duty cycle. 
The seasonal variation in mean duty cycle for UD is more significant than for LE.  As a 
result, the standard deviations, as summarized in  Table~\ref{site_summary}, are also 
higher for both monthly and yearly averages.  In addition, the  mean duty cycle is notably
lower.  Total  observations from UD are available for 60.1\,\% or 3949 days during the period 
of 18 years, the lowest among all sites. The lower coverage from UD was a combination of the
monsoon season and several other long periods of system downtime, for example, a fire broke in
the GONG instrument shelter in August 2010 and the site remained non-operational
until the beginning of 2011. 

\subsection{El Teide, Canary Islands (TD)}

The  El Teide site is located  on the island of Tenerife, in the Canary Islands off
the west coast of Africa, hosted by the Instituto Astrof\'isica de Canarias (IAC). The 
IAC is administered by a public consortium which includes the Universidad de la Laguna.
Similar to other sites, observations from TD are also affected by strong seasonal
variations. Sometimes it is hit by intense dust storm, known as the {\it Calima},  blowing 
off the Sahara Desert. During  the {\it Calima}, strong winds are accompanied by a 
orange haze that significantly reduces the sky transparency and the observations, if any, 
are discarded. Even after being affected by the several months of poor weather conditions 
in Winter and Summer, TD provides a good coverage of about 80.3\,\% of the days.

\begin{figure}    
   \centerline{\includegraphics[width=0.6\textwidth,clip=]{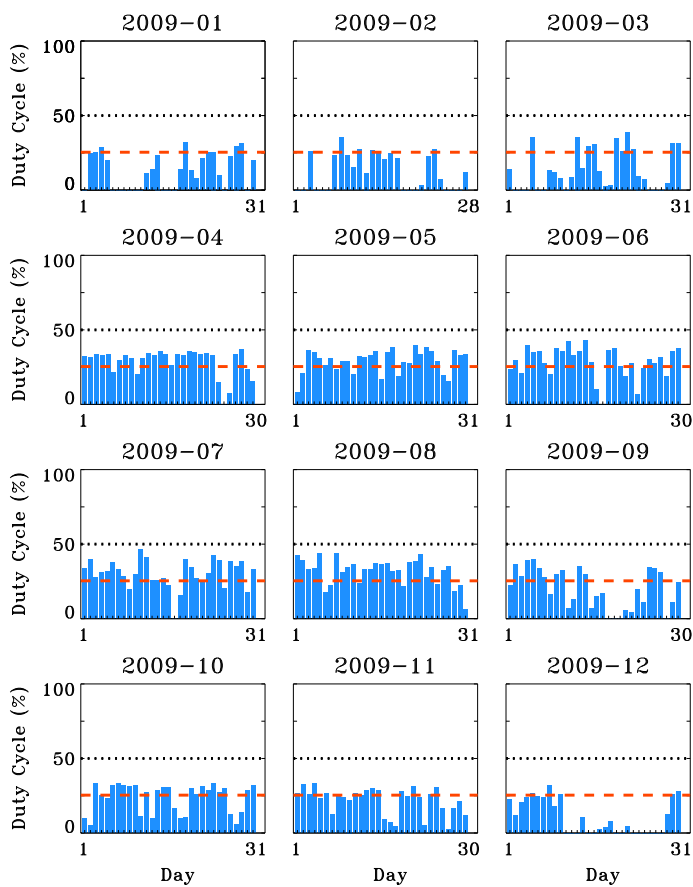}
              }
              \caption{Daily duty cycle for the El Teide (TD) site 
for each month in 2009. The dashed (red) line represents the mean
value over the entire period (see Table~\ref{site_summary}) whereas the dotted (black)
line shows a duty cycle value of 50\,\%.  
   }\label{daily_TD}
   \end{figure}

\begin{figure}    
   \centerline{\includegraphics[width=0.6\textwidth,clip=]{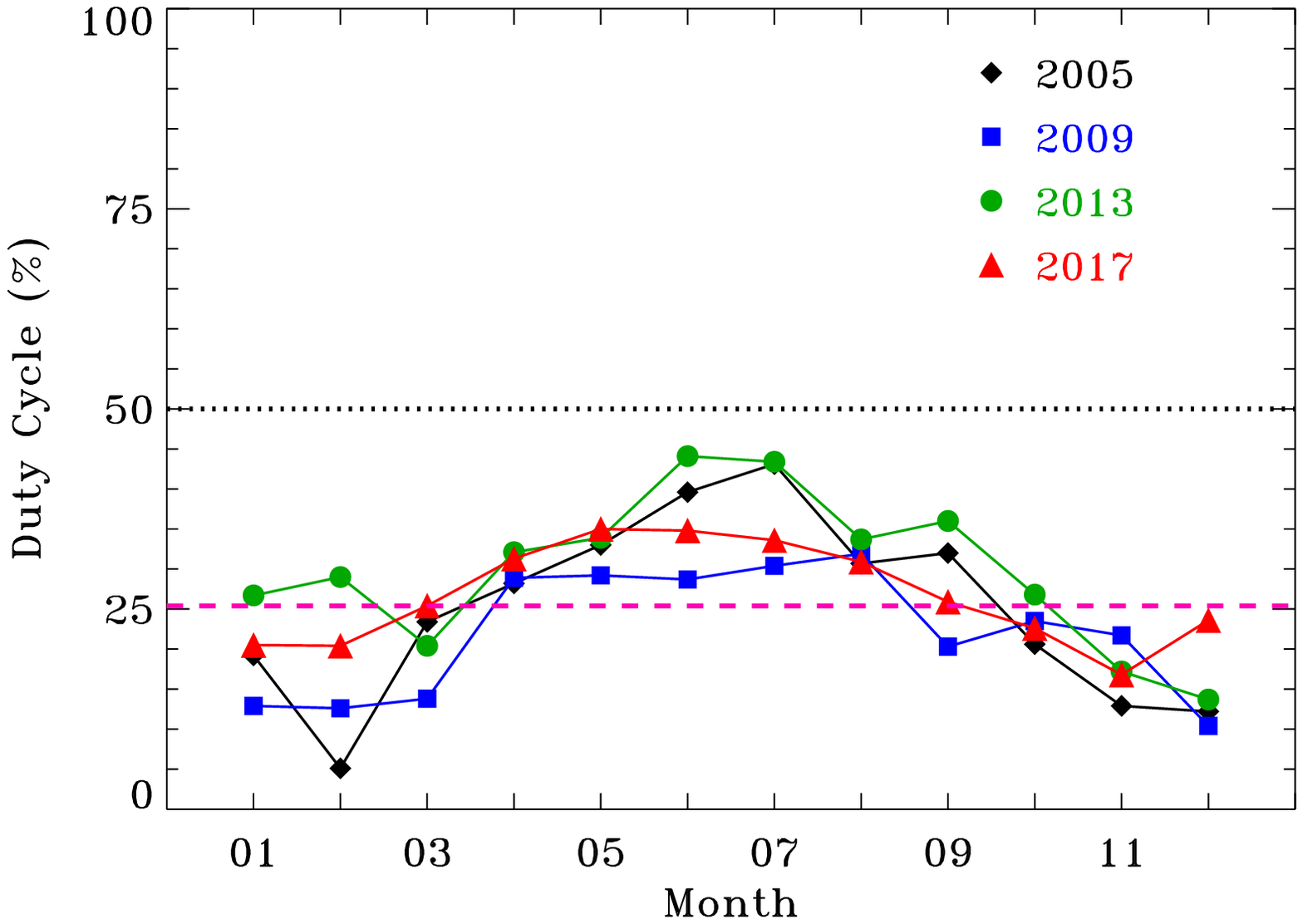}
              }
              \caption{Monthly duty cycle for the Teide (TD) site for
selected years.  Yearly means for 2005, 2009, 2013, and 2017 are 25\,\%, 22\,\%, 30\,\%, 
and 27\,\%, respectively.  The dashed (pink) line represents the mean
value over the entire period (see Table~\ref{site_summary}) whereas the dotted (black)
line shows a duty cycle value of 50\,\%.  
   }\label{monthly_TD}
   \end{figure}

The daily duty cycles for TD are presented in Figure~\ref{daily_TD} for 2009. It is 
evident that daily observations from TD are consistent for several months in Summer, 
thus providing good coverage to the GONG network. This is 
also  apparent from Figure~\ref{monthly_TD} where the duty cycle  from April to September 
for all four years  is consistently higher than other months. From Table~\ref{1site_contri},
one can notice that  TD's contribution to the network duty cycle is about 25\,\% which is 
similar to LE. The independent contribution from the Teide site is 8.7\,\% in the overall 
duty cycle. This site along with LE becomes notably more important when the UD instrument 
is down for the long monsoon season or any other down-time. During a clear day in Summer, 
good observations for almost 12 hours can be obtained from this site. 

Similar to other sites, a strong weather-dependent variation in the TD duty cycle is  apparent 
in Figures~\ref{site_monthly}c  where we show the monthly duty cycle averaged over the 
entire period considered in this study. It is noted that the  duty cycle drops to almost 
one half in the Winter months and the standard deviation in the monthly mean duty cycle is
moderately large with a value of 8.3\,\%. However, the yearly variation is relatively 
small (Figure~\ref{site_yearly}c) with a standard deviation of only 2.6\,\% 
(Table~\ref{site_summary}).

\subsection{Cerro Tololo, Chile (CT)}

\begin{figure}    
   \centerline{\includegraphics[width=0.55\textwidth,clip=]{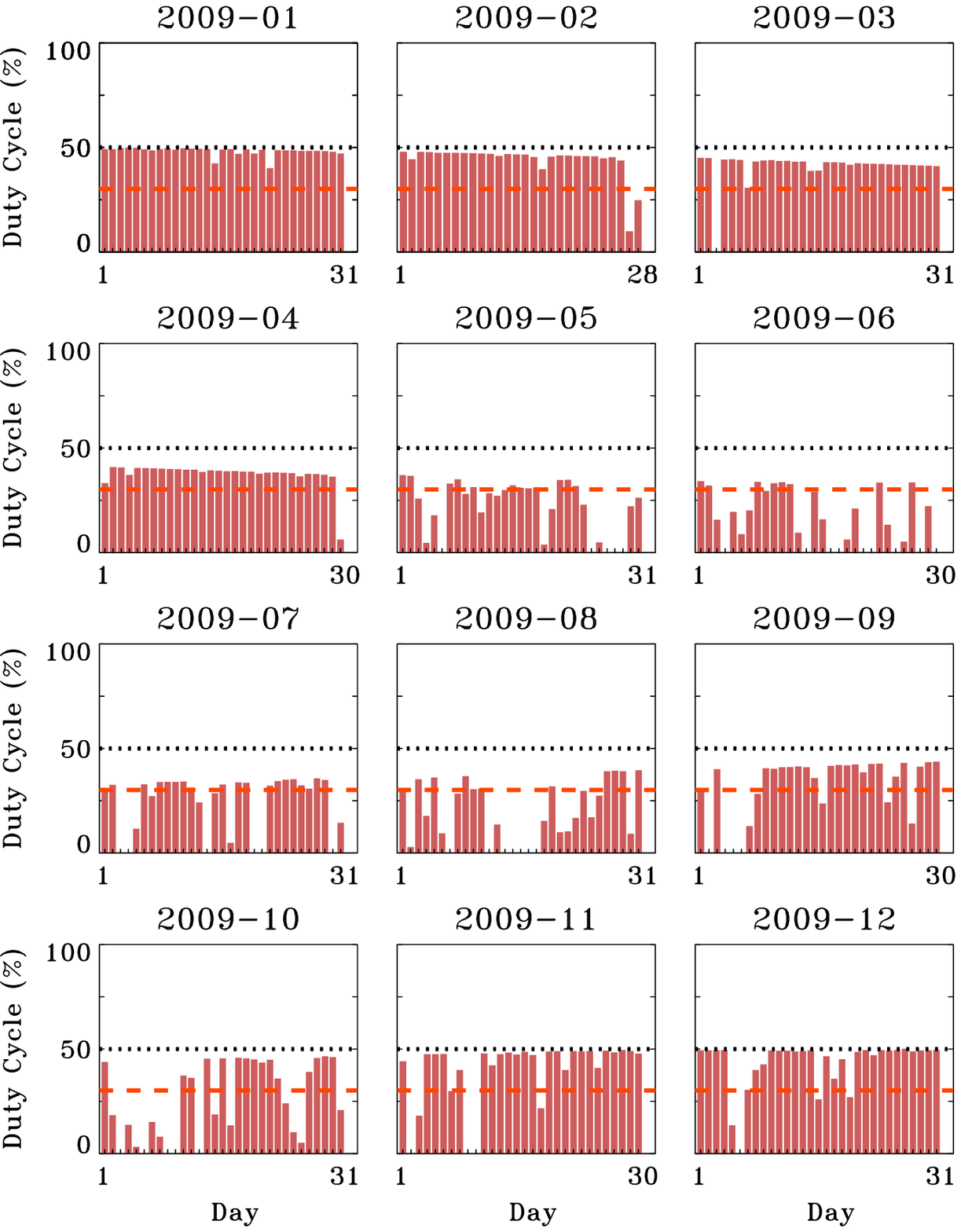}
              }
              \caption{Daily duty cycle for Cerro Tololo (CT) site
for each month in 2009. The dashed (red) line represents the mean
value over the entire period (see Table~\ref{site_summary}) whereas the dotted (black)
line shows a duty cycle value of 50\,\%.  
              }\label{daily_CT}
             \end{figure}
\begin{figure}    
   \centerline{\includegraphics[width=0.65\textwidth,clip=]{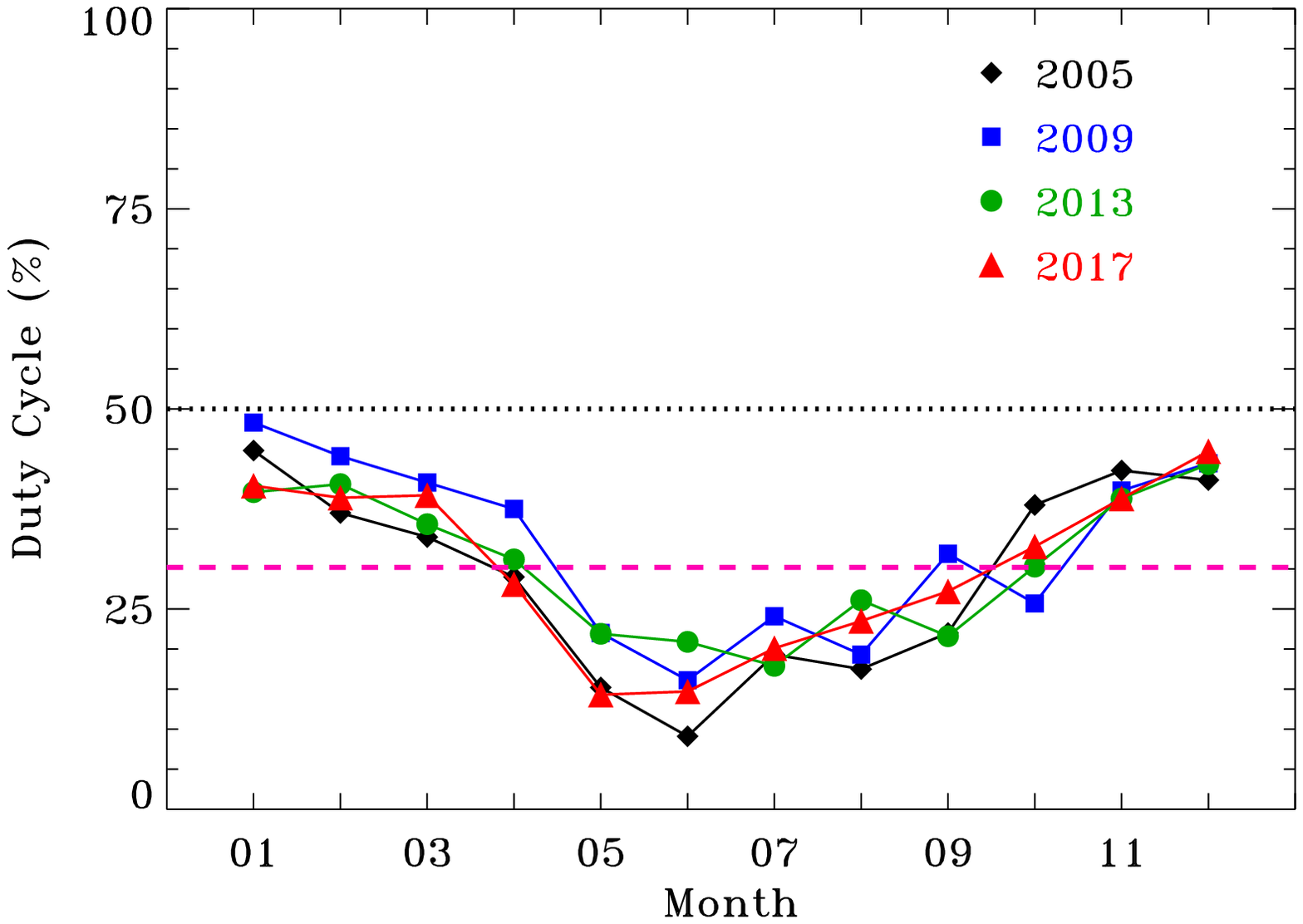}
              }
              \caption{Monthly duty cycle for Cerro Tololo (CT) site for
selected years. Yearly means for 2005, 2009, 2013, and 2017 are 29\,\%, 33\,\%,
31\,\%, and 30\,\%, respectively.The dashed (pink) line represents the mean
value over the entire period (see Table~\ref{site_summary}) whereas the dotted (black)
line shows a duty cycle value of 50\,\%.  
             }\label{monthly_CT}
             \end{figure}

The second GONG site in the southern hemisphere is located on the mountain 
Cerro Tololo in northern Chile. The instrument is located on the grounds of the 
Cerro Tololo Interamerican Observatory (CTIO), a division of NSF’s National 
Optical-Infrared Astronomy Research Laboratory (NSF’s NOIRLab) headquartered in 
Tucson, USA. Due to excellent sky conditions, CTIO is one of the main sites for 
astronomical investigations of the southern skies. This is reflected in the 
number of observations from the individual sites summarized in Table~\ref{1site_contri}. 
CT's contribution to GONG's cumulative observations is the highest and has  
5516 observing days, that is about 84\,\% of the total days. Despite a significantly
large number of observations and about 30\,\% contribution to the network duty cycle,
the independent contribution from CT is only  8.0\,\% while overlapping observations cover 
the remaining 22.1\,\%, as evident from Table~\ref{1site_contri}. This large fraction of 
 overlapping observations are due to the proximity of adjacent sites BB and TD. Moreover, CT 
has the maximum overlap with the neighboring sites' observations. 

Daily variation in duty cycles for the year 2009 is presented in  Figure~\ref{daily_CT}, 
which exhibits stability for several months. Maximum duty cycle of about 50\,\% is
achievable for several months, i.e.  January, February, November, and December. From the
monthly variation for four sample years (Figure~\ref{monthly_CT}), it is clear that the 
number of observations from CT reaches its lowest value in the month of June. This strong
seasonal variations is also noticeable in Figure~\ref{site_monthly}d and the standard 
deviation in the monthly duty cycle is about  9.7\,\%. Aside from the seasonal variation, 
the yearly duty cycles from CT remain reasonably high as shown in  Figure~\ref{site_yearly}d 
and the values remain between 28\,\%\,--\,33\,\% with a standard deviation of 2.2\,\%, 
as given in Table~\ref{site_summary}.

\subsection{Big Bear, USA (BB)}

\begin{figure}    
   \centerline{\includegraphics[width=0.6\textwidth,clip=]{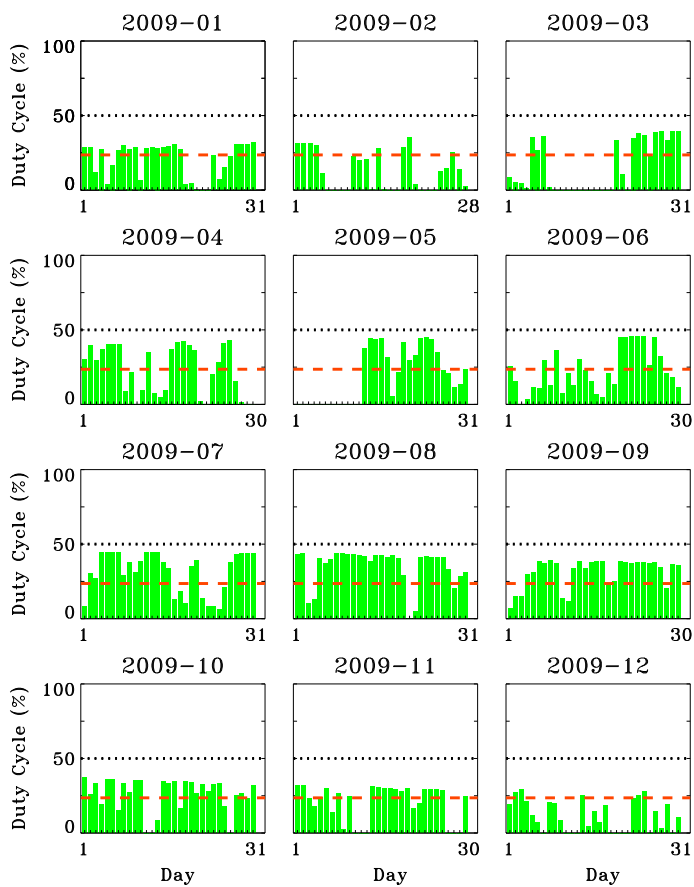}
              }
              \caption{Daily duty cycle for Big Bear (BB) site 
for each month in the year 2009. The dashed (red) line represents the mean
value over the entire period (see Table~\ref{site_summary}) whereas the dotted (black)
line shows a duty cycle value of 50\,\%.  
             }\label{daily_BB}
            \end{figure}
\begin{figure}    
   \centerline{\includegraphics[width=0.6\textwidth,clip=]{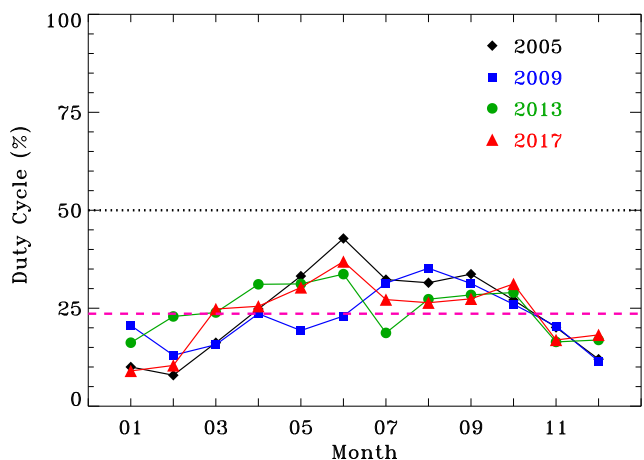}
              }
              \caption{Monthly duty cycle for Big Bear (BB)) site for
selected years. Yearly means for 2005, 2009, 2013, and 2017 are 
24\,\%, 23\,\%, 27\,\%, and 24\,\%, respectively.
 The dashed (pink) line represents the mean
value over the entire period (see Table~\ref{site_summary}) whereas the dotted (black)
line shows a duty cycle value of 50\,\%.  
            }\label{monthly_BB}
            \end{figure}

The only GONG site in the United States mainland is located in the southwest 
on the premises of  Big Bear Solar Observatory (BBSO), managed by the New 
Jersey Institute of Technology (NJIT). The instrument is placed on the north 
shore of an artificial lake in the San Bernardino Mountains in California. 
Similar to other sites, BB also has strong seasonal variation in the duty cycle 
that peaks around June and reaches its lowest value in December\,--\,January.
From the daily duty-cycle histograms, as shown in Figure~\ref{daily_BB}, it is 
evident that the observations are sparse in the Winter months. Although the 
observations from BB are  consistent in Summer months, there are a few years 
when the duty cycle is relatively low and 2009 is among those years.  As 
illustrated in Figure~\ref{monthly_BB}, the monthly duty-cycles for four years 
clearly show  the variation from one year to another. The overall trend is more
pronounced in Figure~\ref{site_monthly}e where the average monthly duty cycle 
is shown. A gradual increase in the duty cycle is observed after February with a 
maximum in June, a slow decrease during July to September and finally, a sharp 
decline is noticed until December. 

We note that the Big Bear site has been very stable throughout the operations of 
GONG and the year-to-year variation is the least among all six GONG sites (See 
Figure~\ref{site_yearly}e). Therefore, the standard deviation in yearly values 
is also a minimum, i.e. only 1.4\,\%,  while the standard deviation in monthly 
values is  7.3\,\%.  Table~\ref{site_summary} shows that the observations 
collected at BB spanned over 5532 days that is about 84.1\,\% of the total 
observing days.  The BB contribution to all GONG observations is around 16.5\,\% 
with three-quarters of them have overlap with other sites as described in
Table~\ref{1site_contri}. Nevertheless, the independent coverage of 4.7\,\% 
in total duty cycle makes this site equally important with other GONG sites.

\subsection{Mauna Loa, USA (ML)}

\begin{figure}    
   \centerline{\includegraphics[width=0.6\textwidth,clip=]{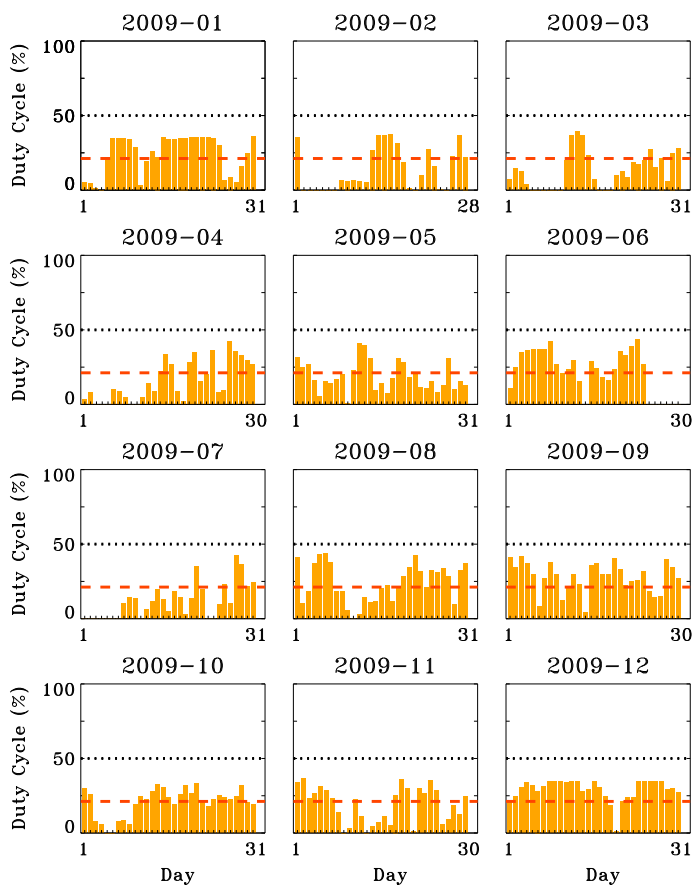}
              }
              \caption{Daily duty cycle for Mauna Loa (ML) site 
for each month in 2009. The dashed (red) line represents the mean
value over the entire period (see Table~\ref{site_summary}) whereas the dotted (black)
line shows a duty cycle value of 50\,\%.  
          }\label{daily_ML}
          \end{figure}

\begin{figure}    
   \centerline{\includegraphics[width=0.6\textwidth,clip=]{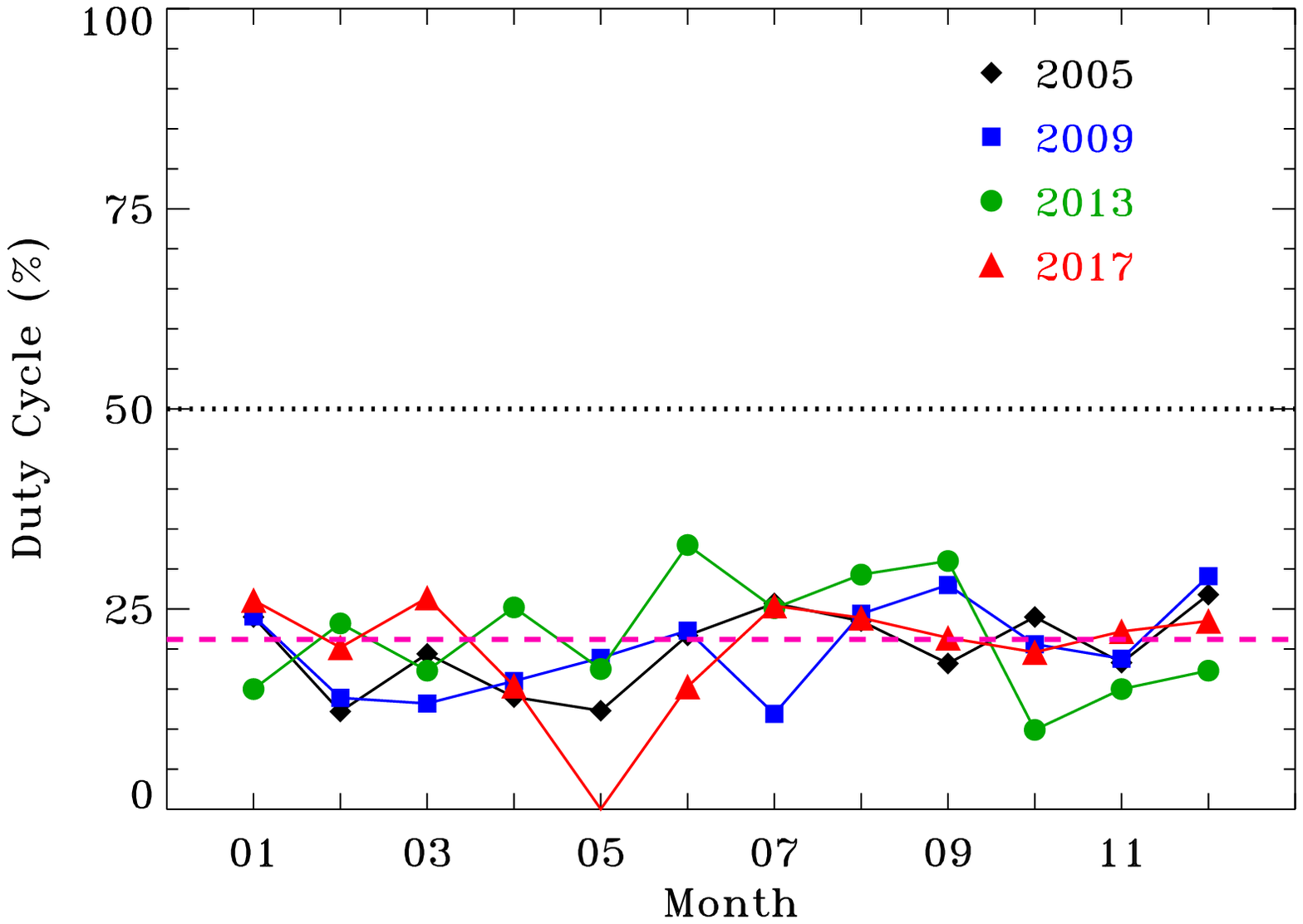}
              }
              \caption{Monthly duty cycle for Mauna Loa (ML) site for 
selected years. Yearly means for 2005, 2009, 2013, and 2017 are 
20\,\%, 20\,\%, 22\,\%, and 20\,\%, respectively. The dashed (pink) line represents the mean
value over the entire period (see Table~\ref{site_summary}) whereas the dotted (black)
line shows a duty cycle value of 50\,\%.  
          }\label{monthly_ML}
           \end{figure}

The sixth GONG site is located at the north side of the Mauna Loa volcano at the Mauna Loa
Observatory (MLO), on the island of Hawai'i (Big Island).  MLO is an atmospheric baseline
station of the Global Monitoring Laboratory (GML) of the National Oceanic and Atmospheric 
Administration (NOAA). Local staff support is provided by the Mauna Loa Solar Observatory 
operated by the High Altitude Observatory, Boulder, Colorado. Unlike other GONG sites, 
Mauna Loa is minimally affected by strong seasonal variations. This is clearly
visible in the daily duty-cycle plot for a year displayed in Figure~\ref{daily_ML}. 
Although we do see variations from month to month, these are not as strong  as in 
other sites and the daily values remain in a similar range throughout the year.  
Further, to evaluate the existence of this pattern in all years, we plot monthly 
values for four individual years in Figure~\ref{monthly_ML}. This shows no distinct 
trend in the monthly variation within a year, however there are random fluctuations, 
but  these do not indicate any strong seasonal pattern. On averaging monthly values 
over the entire period presented in Figure~\ref{site_monthly}f, a mild pattern 
emerges with minor dips around both equinoxes.  Finally, we find that the standard 
deviation in monthly values for ML is only about  2.6\,\% which is the lowest 
among all sites ( Table~\ref{site_summary}). However, the standard deviation in 
yearly values displayed in  Figure~\ref{site_yearly}f is higher than the monthly 
value and also the highest among all sites indicating that ML experienced the 
highest year-to-year variation in the duty cycle.
 
In addition, the number of observing days at ML at 84.2\,\% level is comparable with 
CT and BB,  but the contribution to the total network observations is the second 
lowest (see Table~\ref{1site_contri}).
It is worth pointing out that the independent 
observations from ML cover merely 4.1\,\%  which is the lowest among all six sites; 
more than three-quarters of ML observations overlap with other sites.   

  
\section{Network Duty Cycle}
\label{S-Network}

\begin{figure}    
   \centerline{\includegraphics[width=0.6\textwidth,clip=]{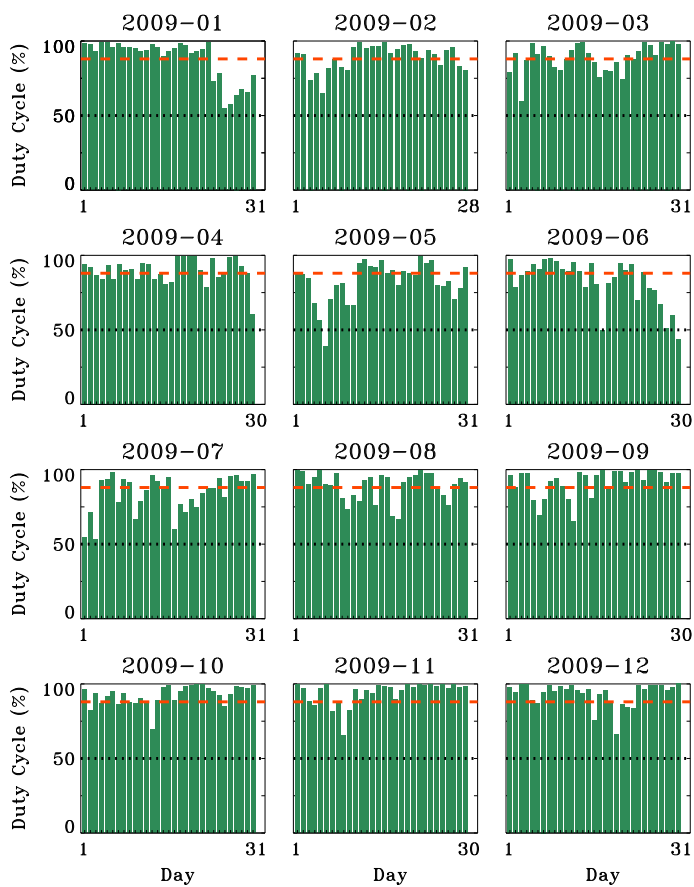}
              }
              \caption{Daily duty cycle of the GONG network for 2009.
             The dashed (red) line represents the mean
value over the entire period (see Table~\ref{nsite}) whereas the dotted (black)
line shows a duty cycle value of 50\,\%.  
              }\label{daily_network}
\end{figure}
\begin{figure}    
   \centerline{\includegraphics[width=0.6\textwidth,clip=]{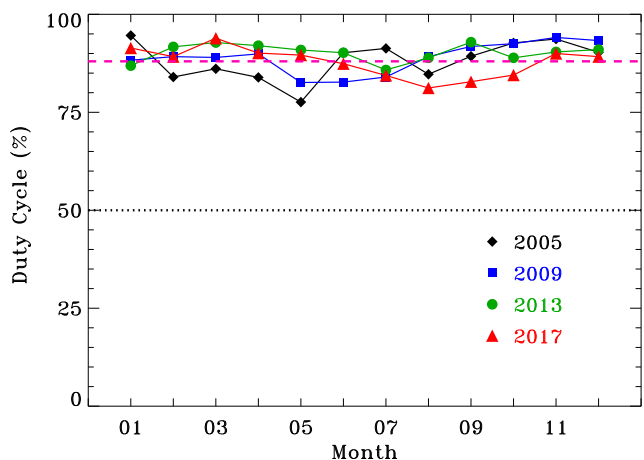}
              }
              \caption{Mean monthly duty cycle of the GONG network for four selected 
years. Yearly means for 2005, 2009, 2013, and 2017 are 88\,\%, 89\,\%, 90\,\%, 
and 88\,\%, respectively. The dashed (pink) line represents the mean
value over the entire period (see Table~\ref{nsite}) whereas the dotted (black)
line shows a duty cycle value of 50\,\%.  
   }\label{monthly_syear}
   \end{figure}
\begin{figure}    
   \centerline{\includegraphics[width=0.5\textwidth,clip=]{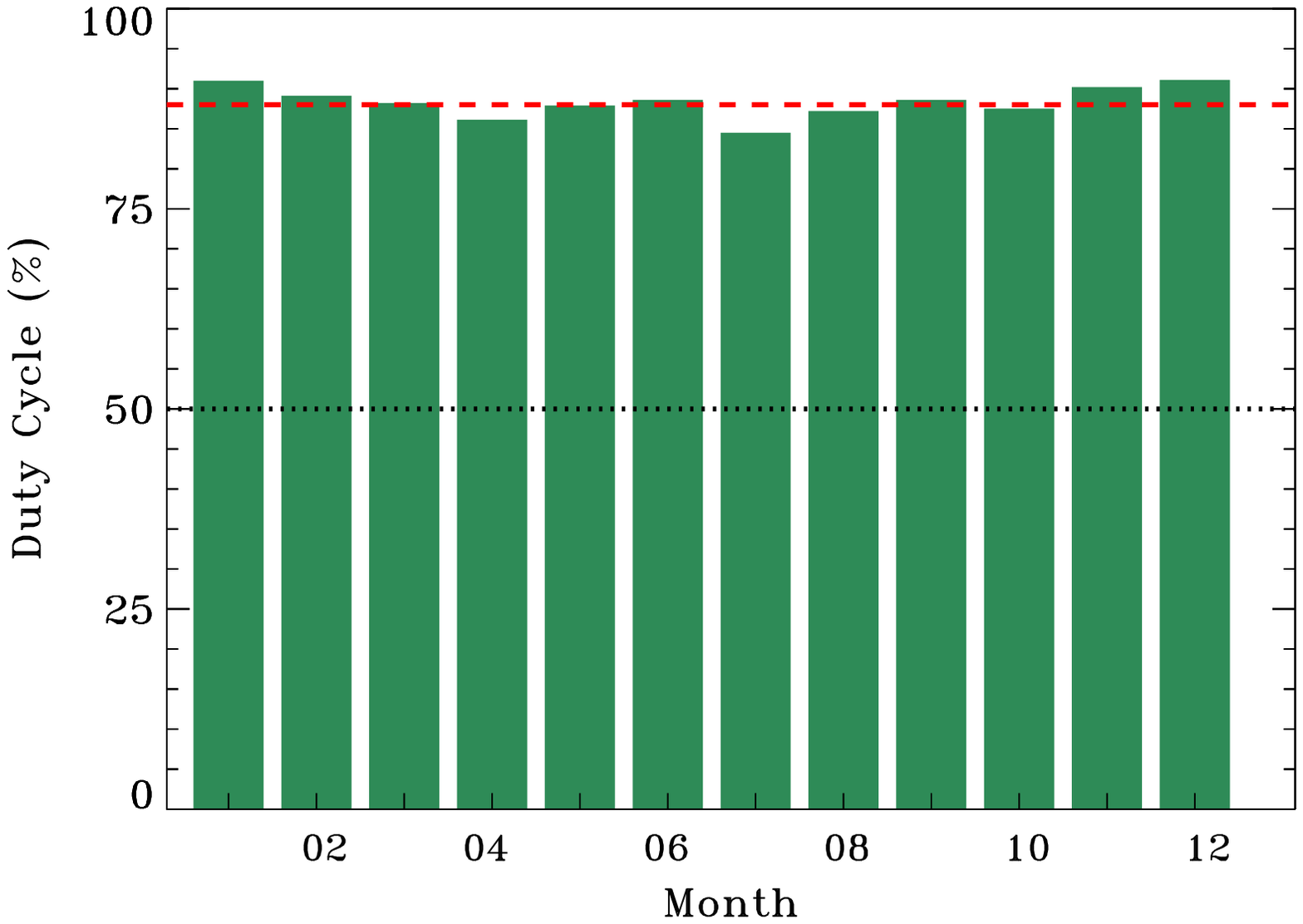}
        \includegraphics[width=0.5\textwidth,clip=]{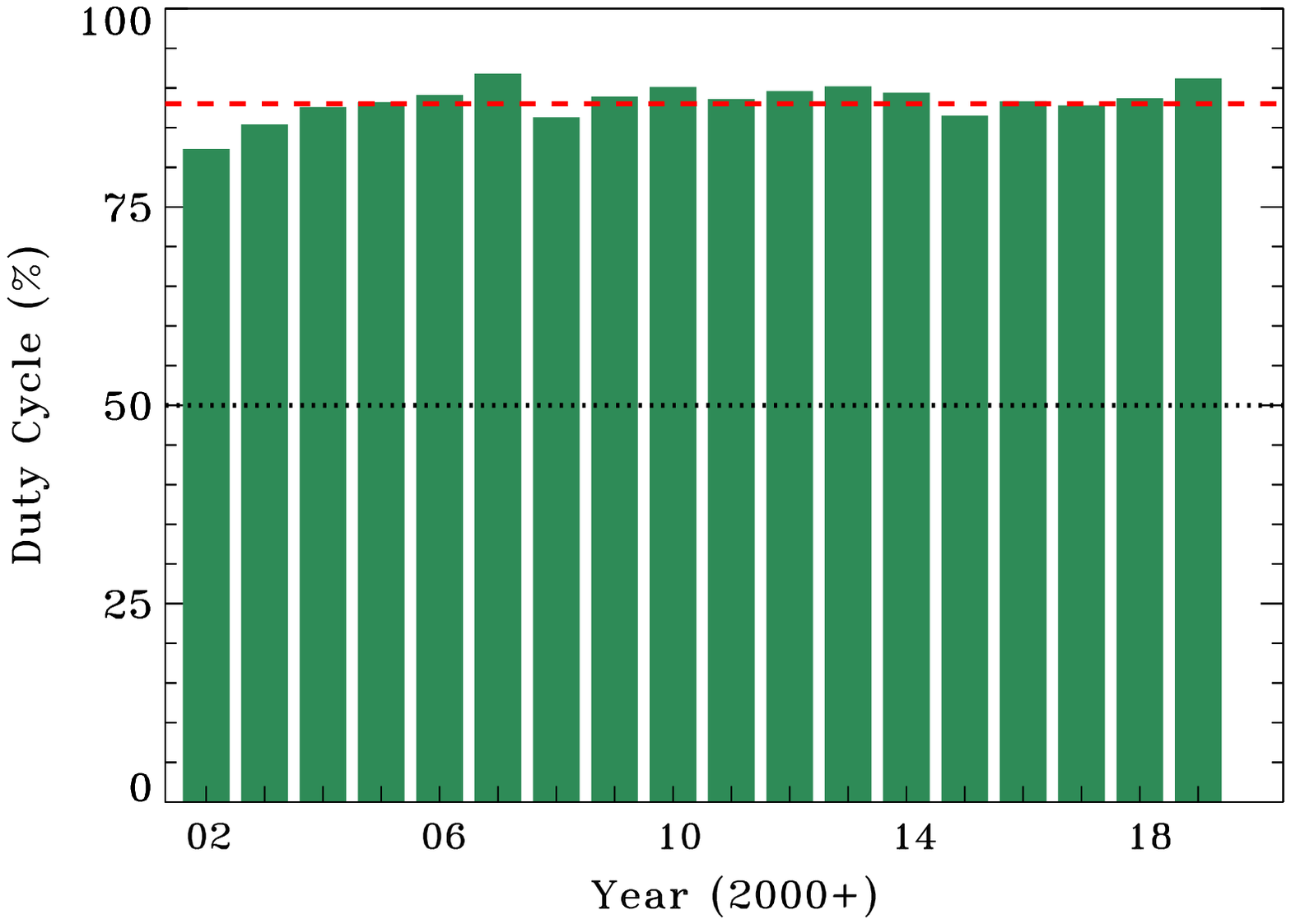}
              }      
              \caption{(Left) Monthly and (Right) yearly variations in the network 
              duty cycles.  The dashed (red) line represents the mean
value over the entire period (see Table~\ref{nsite}) whereas the dotted (black)
line shows a duty cycle value of 50\,\%.  
              }\label{network_monthly_yearly}
  \end{figure}
Section~\ref{S-Sites} clearly demonstrates that all sites except ML are affected 
by strong seasonal patterns. While sites in the northern hemisphere provide good 
coverage during Summer months (humps in June\,--\,July for BB and TD in
Figure~\ref{site_monthly}), this trend reverses in the southern hemisphere (dips in
June\,--\,July for LE and CT). Further, due to the low latitude location, the 
seasonal patterns are less apparent in ML observations.  However, the network
observations after integrating over those from individual sites \cite{Toner03} 
display a consistent coverage throughout the year with significantly high duty 
cycle. Figure~\ref{daily_network} presents network daily duty cycle for a year
exhibiting minimal weather-related patterns and the daily value reaches
$>$ 90\,\% in most cases. We do notice some  days with duty cycles dipping down to 
50\,\% or lower but these are rare. To better understand  the variation, if any,  we 
compute  mean monthly duty cycles for the network for four years and present them in 
Figure~\ref{monthly_syear}.  We do not see any  distinct pattern in all of these
years, which strongly supports the concept of a ground-based network for uninterrupted 
observations with consistent duty cycles.

Finally, we present monthly and yearly duty cycles averaged over the entire period in  
Figure~\ref{network_monthly_yearly}. We notice a mild trend in the monthly duty cycle
(left panel of Figure~\ref{network_monthly_yearly}) throughout the year reaching its 
lowest value in July (84\,\%) and the highest in December/January (91\,\%). We also
notice a mild variation in the yearly values.  The yearly mean is relatively low in 
the first two years and then remains around the mean over the entire period for rest 
of the years.  The low standard deviations of 1.9\,\% and 2.2\,\% in monthly and 
yearly values, respectively  exhibit excellent stability in the network duty cycle,
even after more than 25 years of its operations. In addition, Figure~\ref{site_yearly}
also indicates that there are no long-term changes in the duty cycles of any of the 
six network sites though the contribution from each site is affected by the local weather 
as well as the instrument down time.  The lack of long-term variations in duty cycles 
further implies that there are no significant changes in climate in any of the GONG 
network sites. These findings clearly demonstrate that a ground-based network can 
reliably study long term solar variability. 

\begin{figure}    
   \centerline{\includegraphics[width=0.65\textwidth,clip=]{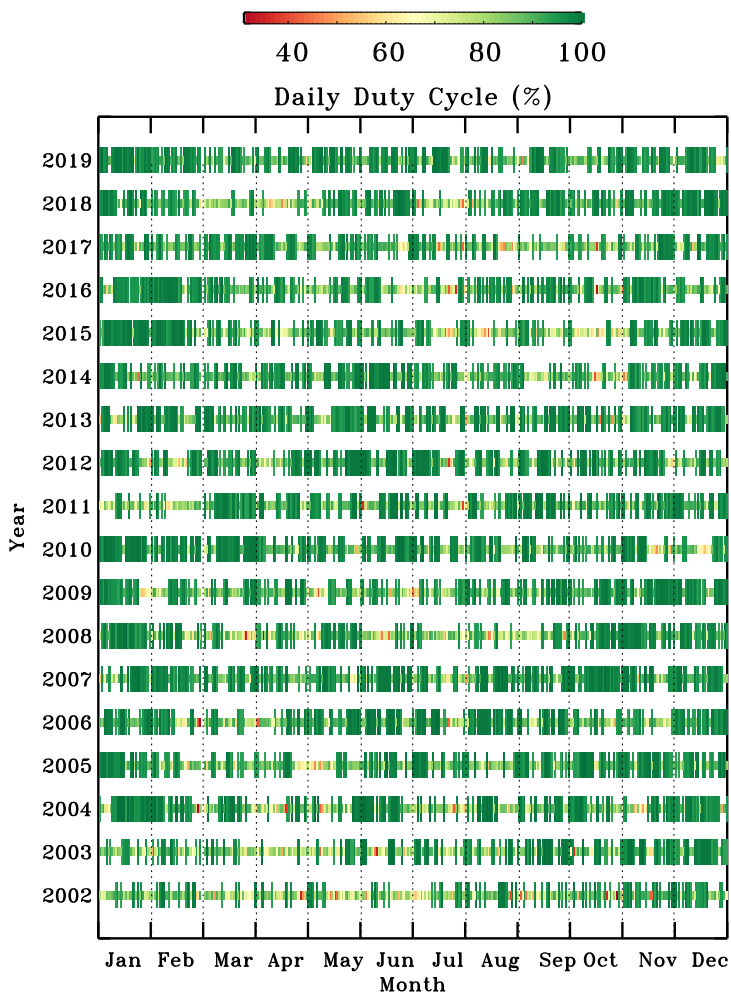}
              }
              \caption{Daily duty cycle of GONG network for the entire period covered in
this study, i.e. 2002\,--\,2019. Smaller bars represent days with duty cycle 
lower than the median value, i.e. 92\,\%.
               }\label{duty_daily_all}
\end{figure}

\begin{figure}    
   \centerline{\includegraphics[width=0.65\textwidth,clip=]{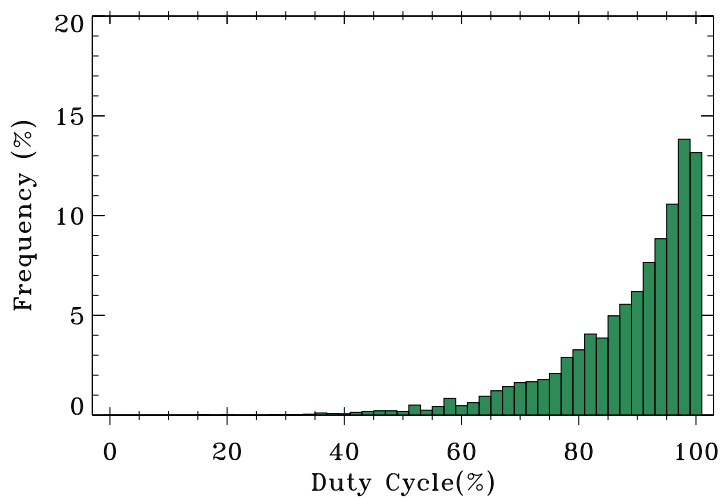}
              }
              \caption{Histogram of daily duty cycle of the fully calibrated
              images for the entire period.
              }\label{daily_histo_all}
\end{figure}

The mean duty cycle over 18 years is 88.3\,\% and the median value  is 92.0\,\%.
Figure~\ref{duty_daily_all} displays the duty cycle corresponding to each day for the 
entire analysis period; smaller bars representing days where the duty cycle is lower 
than the median values.  It is also observed  that most days with lower duty cycles 
are around mid-year which is also apparent in monthly mean values displayed in 
Figure~\ref{network_monthly_yearly}. Distribution of the daily values for bin-size 
of 2\,\% is given in Figure~\ref{daily_histo_all}. We find that  only 82 days out
of 6574 days covered in this study have the duty cycle less than 50\,\%.
Moreover, almost 13\,\% of the total days, i.e. 865 days, have a duty cycle of 99\,\%
or greater of which 106 days have no gaps, i.e., 100\,\% duty cycle. In general,  
the frequency of occurrence  gradually decreases with decreasing duty cycle except 
when values lie between 97\,\%\,--\,98\,\% where we notice a slight increase.

\begin{figure}    
   \centerline{\includegraphics[width=0.65\textwidth,clip=]{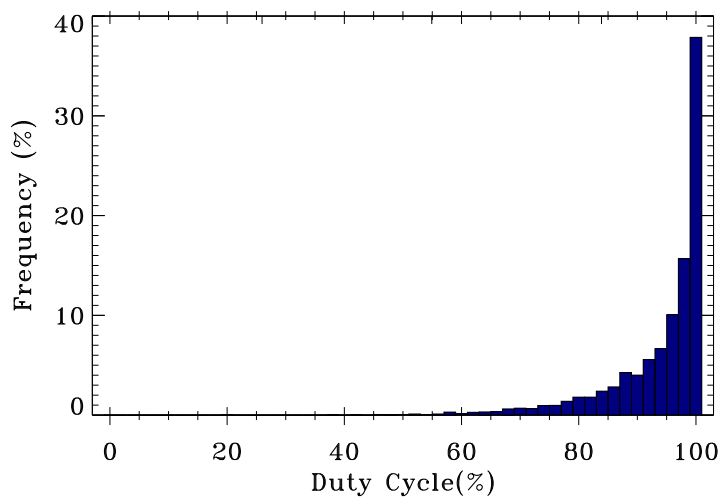}
              }
              \caption{Distribution of daily duty cycle computed from all the images 
              taken at sites. Please note that these images are  partially calibrated 
              (see text for explanation).   
              }\label{daily_histo_NRT}
\end{figure}

It is important to mention that the above analysis is based on the {\it fully calibrated }
images where each and every individual site image is passed through the pre-defined 
quality control checks \cite{Clark04}. During this process, some images are rejected 
every day affecting the site as well as network duty cycles. To estimate  the impact 
of the rejection criteria on the network duty cycle, we display the distribution of 
the duty cycle computed from the partially calibrated site images in
Figure~\ref{daily_histo_NRT}. 
Note that these partially calibrated images do not pass through all quality checks
and may be counted for initial observations. A comparison between the duty cycles for
partially and fully calibrated data reveals that the mean duty cycle obtained from 
the partially calibrated images is higher with 
the mean value of 93.5\,\% and the median value of 97\,\%. This is in excellent 
agreement with the GONG site-survey study where \cite{Hill94b} suggested an 
achievable duty cycle value of 93.3\,\%. This also indicates  that the quality control
matrices reduce the duty cycle by about 5\,\%  on an average, however there are
large number of days when the duty cycle is not affected in this process, primarily 
due to multiple-site observations for a given minute. Compared to fully calibrated 
data where only 13\,\% of the days have duty cycle 99\,\% or more, the partially
calibrated data have about 40\,\% of the  days with duty cycle in this range. This
reduction in duty cycle is also seen in another six-site network, BiSON, where the 
duty cycle of 82\,\% reduces to 78\,\%  when fully calibrated images were used
\cite{Hale16}.

\subsection{Comparison with Previous Results}
\label{S-Comp}

\begin{figure}    
   \centerline{\includegraphics[width=0.65\textwidth,clip=]{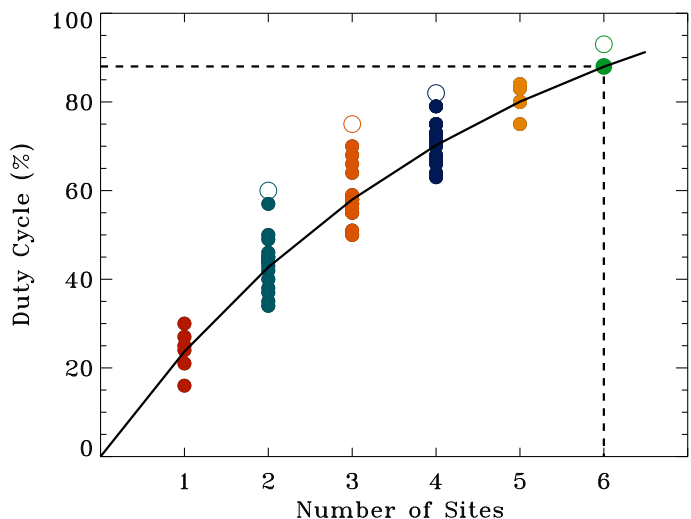}
              }
              \caption{Dependence of the duty cycle on the number of sites. Filled circles
             represent values for each number of stations with different GONG-site 
             combinations. Open circles are for the annual mean duty cycle computed using 
             hypothetical networks as presented by \cite{Hill85}.
             Solid line represents fit to the data with a exponential function described
             in the text. 
             }\label{fig_summary}
\end{figure}
Based on the hypothetical networks, \citename{Hill85} \citeyear{Hill85} reported 
that two-, three-, four-, and six-site networks might achieve 60\,\%, 75\,\%, 82\,\%,
and 93\,\% annual mean duty cycle. To check the feasibility of these duty cycle values,
we also computed duty cycles for all possible speculative networks using GONG's  fully
calibrated site images and the results are displayed in Figure~\ref{fig_summary}. The
maximum  duty cycles for two-, three-, four-, five-, and six-site combinations are
57\,\%, 70\,\%, 79\,\%, 84\,\%, and 88\,\%, respectively. These values are smaller by 
a few percent than those reported in  \citename{Hill85} \citeyear{Hill85}. These are
shown by the open symbols.  It should be noted that the duty-cycle values obtained in
this study are also affected  by the downtime of each site due to instrument repairs
and preventive maintenance trips which were not considered in the simulated analysis.
However, as discussed earlier, the use of fully calibrated images reduces  the duty
cycle  by about 5\,\%  and  thus the values obtained in this study are comparable with
those reported by \cite{Hill94b}.

To study the impact of number of sites on the duty cycle, we finally fit duty cycle
values obtained for different site combinations to a function. The best fit is found
for an exponential function of the form of $f(n)$ = -1.20\,e$^{-0.22n} +1.20 $, 
where $f$ is the duty cycle, and  $n$ is the number of sites
(Figure~\ref{fig_summary}).  It
clearly displays that the duty cycle increases exponentially with the increasing number
of sites, however the locations of  the observing sites play a critical role in
achieving desired duty cycle. We also find that adding another site to the six-site network 
will improve the duty cycle notably.

\section{Simultaneous Observations from Multiple Sites}
\label{S-Simul}

\begin{table}
\caption{\label{site_simul} Multi-site simultaneous contribution to the network  duty cycle. }
\begin{indented}
\item[]\begin{tabular}{@{}ccrl}     
  \br                  
Simultaneous         & Count & Duty Cycle & Details\\
 Observations Type                    &[Minutes]&             \\
\mr
Total                                      & 4.24 $\times$ 10$^{6}$  &  44.84\,\% \\
2 sites & 3.37 $\times$ 10$^{6}$  &  35.60\,\% & Table 6\\
3 sites & 8.20 $\times$ 10$^{5}$  &  8.66\,\% & Table 7 \\
4 sites & 5.45 $\times$ 10$^{4}$  &  0.58\,\% & Table 8\\
  \mr
\end{tabular}

\end{indented}
\end{table}
In a well-organized network, it is important that simultaneous observations are 
carried out at adjacent sites. This will not only minimize interruptions in 
observations but also helps in a better data calibration. In ideal conditions, 
the GONG network should be providing overlapping observations all of the 
time. However as presented in Table~\ref{nsite}, almost 11\,\% of the 
period considered in this study
was not covered by any site and the rest of the coverage was equally divided  among 
single-site and overlapping observations. Table~\ref{site_simul} illustrates the 
details of simultaneous observations from two, three, and four sites.  As expected, 
the maximum simultaneous coverage comes from two-site overlap. This is in agreement 
with the site-survey study by \cite{Hill94b} where they also achieved a maximum 
overlap of 44.8\,\% between two sites. Further, the single-site observations were 
limited to 28.9\,\%, which is lower than this study, and three- and four-site
simultaneous observations contributed to 18.5\,\% and 1.7\,\% of the total observing time.  

\subsection{Simultaneous Observations from Two Sites}
\label{S-2Simul}
\begin{table}
\caption{\label{2site_simul} Simultaneous coverage from two sites in the network.  }
\begin{indented}
\item[]\begin{tabular}{@{}llcccccc}     
  \br                  
Site 1 & Site 2  & \multicolumn{2}{c} {Total Coverage} && \multicolumn{2}{c} {Simultaneous Coverage} \\
\cline{3-4} \cline{6-7}
       &         & [Minutes]         & Duty Cycle &&[Minutes]  & Duty Cycle \\
\mr
  LE  & UD &  3.2 $\times$ 10$^{6}$  & 34.4\,\%  &&  6.9 $\times$ 10$^{5}$  &  7.3\,\%  \\
      & TD &  4.8 $\times$ 10$^{6}$  & 50.5\,\% & &  9.7 $\times$ 10$^{4}$  &  1.0\,\%  \\
      & CT &  5.4 $\times$ 10$^{6}$  & 56.8\,\%  &&  211                    & $<$0.1\,\% \\
      & BB &  4.6 $\times$ 10$^{6}$  & 49.0\,\%  &&  6.9 $\times$ 10$^{4}$  &  0.7\,\%  \\
      & ML &  4.1 $\times$ 10$^{6}$  & 43.8\,\% & &  3.1 $\times$ 10$^{5}$  &   3.3\,\% \\ 
\\
  UD  & TD &  3.6 $\times$ 10$^{6}$  & 38.4\,\% & &  2.2 $\times$ 10$^{5}$  &   2.3\,\% \\
      & CT &  4.3 $\times$ 10$^{6}$  & 45.9\,\% & &  1.0$\times$ 10$^{4}$   &   0.1\,\% \\
      & BB &  3.7 $\times$ 10$^{6}$  & 39.5\,\% & &  34                     & $<$0.1\,\% \\
      & ML &  3.5 $\times$ 10$^{6}$  & 36.7\,\%  &&  1.7 $\times$ 10$^{4}$  & 0.1\,\%   \\
\\ 
TD    & CT &  4.2 $\times$ 10$^{6}$  & 44.8\,\%  &&  6.9 $\times$ 10$^{5}$  & 7.3\,\%   \\
      & BB &  4.2 $\times$ 10$^{6}$  & 44.0\,\%  &&  1.4 $\times$ 10$^{5}$  &  1.5\,\%  \\
      & ML &  4.3 $\times$ 10$^{6}$  & 45.2\,\%  && 1.6 $\times$ 10$^{4}$   &   0.1\,\%  \\
\\    
 CT   & BB &  4.0 $\times$ 10$^{6}$  & 42.4\,\%  && 4.0 $\times$ 10$^{5}$   & 4.2\,\%  \\
      & ML &  4.1 $\times$ 10$^{6}$  & 43.5\,\%  && 2.8 $\times$ 10$^{5}$   & 3.0\,\%   \\
\\ 
BB    & ML &  3.3 $\times$ 10$^{6}$  & 34.9\,\% & &  4.7 $\times$ 10$^{5}$  &  4.4\,\%  \\
\mr
\end{tabular}
\end{indented}
\end{table}

Table~\ref{site_simul} shows that the two-site  simultaneous observations provide 
coverage of about 35.6\,\%  over the entire period. We present in Table~\ref{2site_simul}  
the details of the two-site observations and their coverage. In addition,  every site 
has some fraction of overlapping observations with adjacent sites. It can be easily 
seen that two-site observations provide coverage between 34\,\%\,--\,56\,\%, however 
the simultaneous coverage ranges between $<$0.1\,\% and  7.3\,\%. It is obvious that the 
simultaneous coverage depends on the  locations of the sites in the network.  The 
maximum overlap is observed between two pairs of sites, i.e.,  LE and UD, and TD and CT,  
with a contribution of 7.3\,\% to the total duty cycle. However, LE and UD have
more overlapping observations, i.e., 21.6\,\% of their total coverage compared to TD 
and CT with an overlapping observation of 16.4\,\%.

\subsection{Simultaneous Observations from Three Sites}
\label{S-3Simul}

\begin{table}
\caption{\label{3site_simul} Simultaneous coverage from three sites in the network. }
\begin{indented}
\item[]\begin{tabular}{@{}lllccccccc}     
  \br          
Site 1 & Site 2  & Site 3 &\multicolumn{2}{c} {Total Coverage}  && \multicolumn{2}{c} {Simultaneous Coverage}   \\
\cline{4-5} \cline{7-8}
       &         &        & [Minutes]       & Duty Cycle    &               & [Minutes]  & Duty Cycle \\
\mr
   LE  &  UD     &  TD    &  5.3 $\times$ 10$^{6}$ &  55.9\,\% &&   5.4 $\times$ 10$^{4}$ &  0.6\,\%    \\   
   LE  &  UD     &  ML    &  4.9 $\times$ 10$^{6}$ &  51.4\,\% &&   2.7 $\times$ 10$^{4}$ &  0.3\,\% \\
   LE  &  CT     &  BB    &  6.4 $\times$ 10$^{6}$ &  67.8\,\% &&   2.0 $\times$ 10$^{2}$ & $<$0.1\,\% \\   
   LE  &  CT     &  ML    &  6.3 $\times$ 10$^{6}$ &  66.0\,\% & &  3.9 $\times$ 10$^{2}$  &  $<$0.1\,\% \\   
   LE  &  BB     &  ML    &  5.4 $\times$ 10$^{6}$ &  56.7\,\% & &  4.4  $\times$ 10$^{4}$ &  0.5\,\%    \\
\\
   UD  &  TD     &  CT    &  5.5 $\times$ 10$^{6}$ &  57.7\,\% &&   1.2  $\times$ 10$^{4}$ &  0.1\,\%    \\
\\
   TD  &  CT     &  BB    &  5.2 $\times$ 10$^{6}$ &  55.1\,\% & &  2.4 $\times$ 10$^{5}$  &  2.5\,\%    \\
   TD  &  CT     &  ML    &  5.5 $\times$ 10$^{6}$ &  57.6\,\% & &  2.5  $\times$ 10$^{4}$ &  0.3\,\%    \\
   TD  &  BB     &  ML    &  5.2 $\times$ 10$^{6}$ &  54.9\,\% & &  3.9 $\times$ 10$^{4}$ &  0.4\,\%    \\
\\
   CT  &  BB     &  ML    &  4.8 $\times$ 10$^{6}$ &  50.4\,\% & &  3.8 $\times$ 10$^{5}$  &  4.0\,\%    \\
\mr
\end{tabular}

\end{indented}
\end{table}

Compared to two-site observations,  three-site simultaneous coverage is less frequent 
and the contribution is about 8.66\,\% in the total duty cycle.  Table~\ref{3site_simul}
presents the coverage between three sites where  we have included only those sites 
that observed simultaneously at any point in the entire period.  We find only two 
three-site combinations that observed simultaneously for  $>$1\,\% in the total
coverage. Among these, the maximum overlap is obtained between  CT, BB, and ML.

\subsection{Simultaneous Observations from Four Sites}
\label{S-4Simul}

For a tiny fraction in network duty cycle, i.e. $<$1\,\% ,  four GONG sites observed 
simultaneously. As presented in Table~\ref{4site_simul}, the maximum contribution
came from TD, CT, BB, and ML. These sites covered 63\,\% of the available time  
together in 6574 days of which only 0.57\,\%, i.e. 54,423 min observed simultaneously.  
There are two other site combinations in this category, however their contribution
to the network duty cycle is less than 0.01\,\%.

\begin{table}
\caption{\label{4site_simul} Simultaneous coverage from four sites in the network. }
\begin{indented}
\item[]\begin{tabular}{@{}lccccccccc}     
  \br         
        Sites  & \multicolumn{2}{c} {Total Coverage}  && \multicolumn{2}{c} {Simultaneous
        Coverage}   \\
\cline{2-3} \cline{5-6}
                    &  [Minutes]         & Duty Cycle      &  & [Minutes]  & Duty Cycle \\
\mr
LE, CT, BB, ML & 6.8 $\times$ 10$^{6}$ &   72.3\,\,\%   &&   328        &     $<$0.01\,\,\%  \\
LE, UD, BB, ML & 6.1 $\times$ 10$^{6}$ &   64.3\,\,\%   & &  14         &$ <$0.01\,\,\%\\
TD, CT, BB, ML & 6.0 $\times$ 10$^{6}$ &   63.0\,\,\%   &&   54423      &  0.57\,\%           \\
\mr
\end{tabular}

\end{indented}
\end{table}

\section{Significance of High Duty Cycle}
\label{S-Impact}
GONG's uninterrupted full-disk observations have been widely used in numerous 
science projects as well as space-weather operations/monitoring where duty cycle
plays a pivotal role.   In particular, helioseismic studies have been greatly
benefited from these long span of observations. For example, the continuous data
with consistently high duty cycle have allowed the study of temporal variability
in global \cite{Jain09,Jain11,Broomhall17} as well as local oscillation mode 
characteristics \cite{Tripathy13,Tripathy15} and their connection with the  
11-year cyclic behaviour of the surface magnetic activity. In addition, these 
data have also revealed other periodicities in helioseismic data, e.g.,  the 
quasi-biennial oscillations (QBO) in oscillation frequencies \cite{Simoniello13a}. 
Further, long consistent helioseismic data  are also crucial in constraining the 
fundamental stellar properties, particularly in asteroseismology where most 
stellar observations cover relatively short time intervals \cite{Howe20}.

\begin{figure}    
\hskip0.35in
   \centerline{\includegraphics[width=0.46\textwidth,clip=]{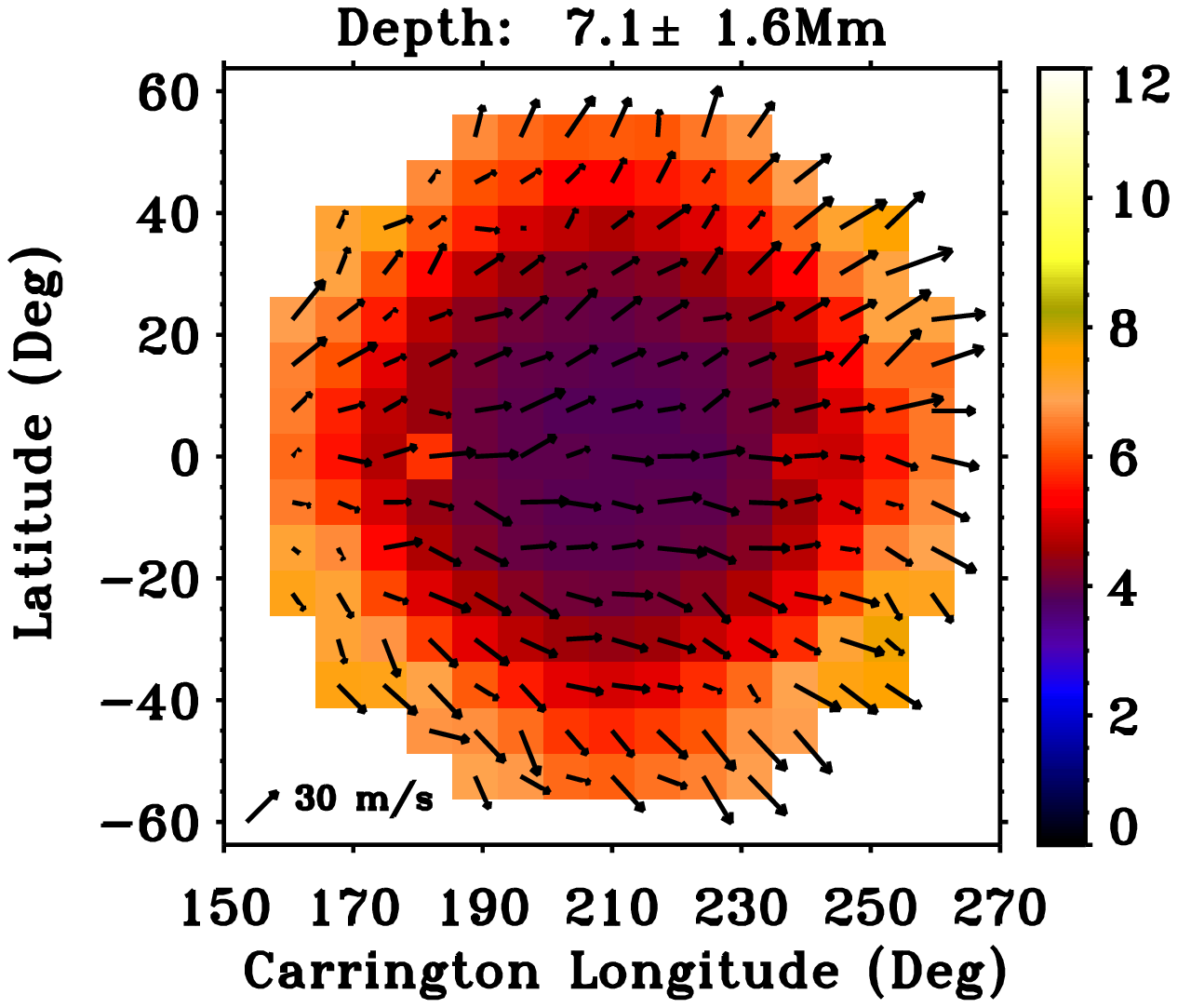}
   \hskip-0.98in
   \includegraphics[width=0.46\textwidth,clip=]{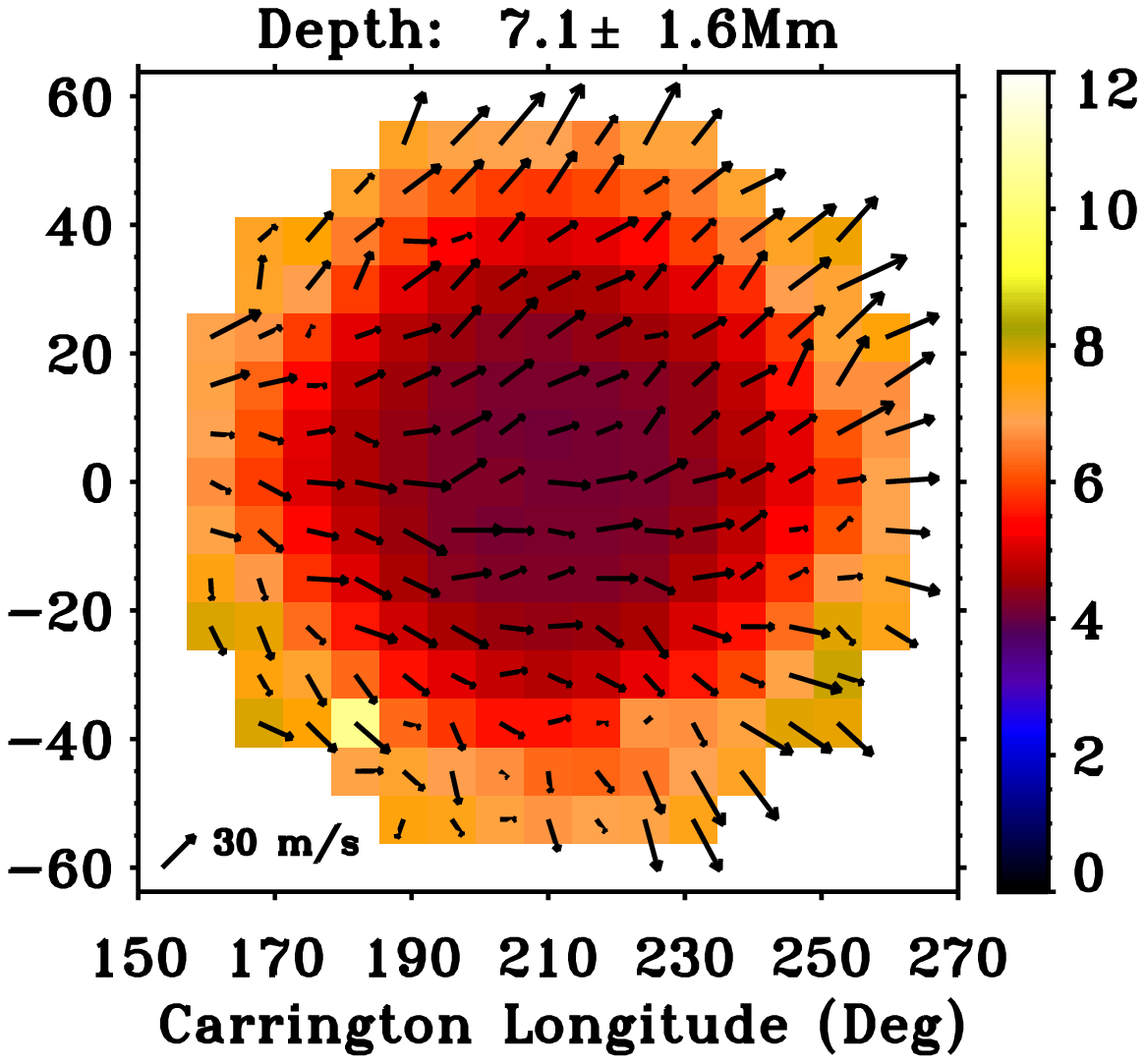}
   \hskip-0.98in
    \includegraphics[width=0.46\textwidth,clip=]{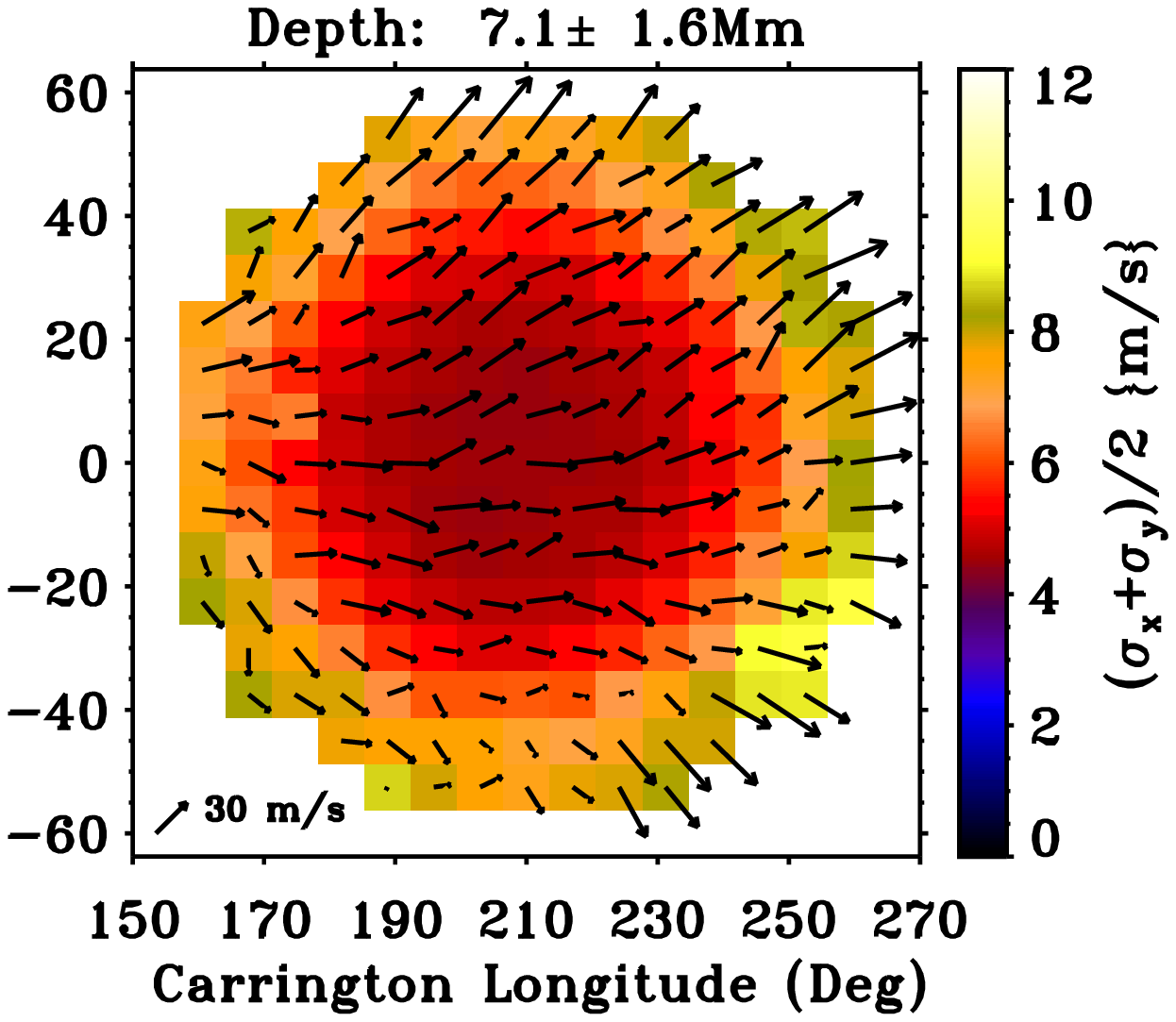}
      }        
              \caption{Effect of duty cycle on the subsurface flow measurements computed using 1664
              min GONG Dopplergrams with (Left) 100\,\%, (Middle) 79\,\%,and 66\,\% (Right) duty 
              cycles. Arrows show the flow magnitude and the direction, and the background depict
              the uncertainties in these estimates.
             }\label{flows}
             
\end{figure}

Another benefit of Doppler observations with high duty cycle is the reliable
inferences of subsurface flows which are crucial in understanding the plasma motion 
below the surface and the solar dynamo. Several studies have been carried out where 
torsional oscillation patterns \cite{Howe00,Howe18b}, subsurface zonal and 
meridional flows \cite{Komm18,Basu19,Gizon20}, and deep meridional flows 
\cite{Jackiewicz15} have been reliably inferred. To emphasize the importance of 
high-duty cycle in helioseismic studies, we compare horizontal flows (Figure~\ref{flows}) 
computed utilizing the technique of ring diagrams \cite{Hill88} where  1664 full-disk 
Doppler images taken at 1 min cadence are used. In the figure, arrows show the flow 
magnitude and the direction,  and the background depict the uncertainties in these 
estimates.  The left panel shows the horizontal flows with a 100\% duty cycle while 
the middle and right panels display the flows measured with reduced duty cycles  of 79\% 
and 66\%, respectively.  Lower duty cycles are computed by truncating observations from  
100\,\% duty cycle data. From a comparison of these three panels, we notice that the 
flow magnitudes as well as the errors increase with the decrease of the duty cycles. 
Thus, the reliable flow measurements require a higher duty cycle. 

\begin{figure}    
   \centerline{\includegraphics[width=0.65\textwidth,clip=]{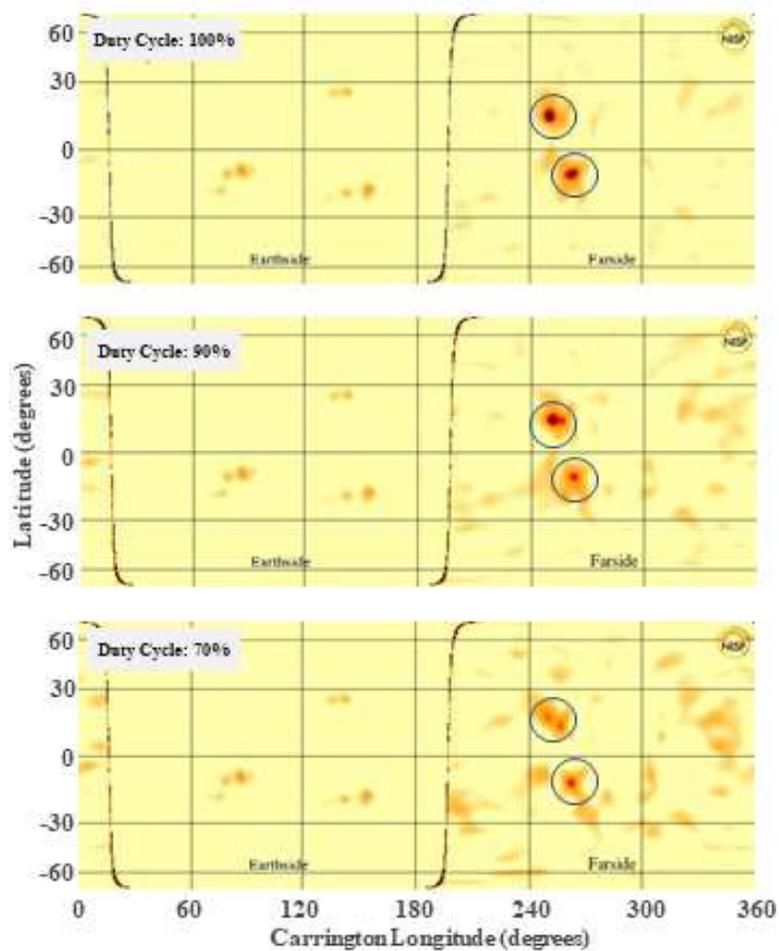}
              }
              \caption{Effect of duty cycle on the farside helioseismic maps computed
              using 24 h GONG Dopplergrams with (top) 100\,\%, (Middle) 90\,\%, and (Bottom) 70\,\%
              duty cycles.  Vertical lines show the boundaries between farside and nearside 
              (direct observations) hemispheres. Regions of high magnetic field, i.e.active 
              regions, are shown by the darker regions on both sides. Identified active regions 
              on the farside are marked with circles. It is clearly visible that the low duty 
              cycle increases noise in the farside maps and also reduces the signal to noise.
             }\label{farside}
\end{figure}

 GONG has also become an important data source for uninterrupted high-cadence
magnetic-field and H$\alpha$ observations, primarily due to significantly 
high duty cycle. These 
are being used in various space-weather  forecast systems \cite{Petrie08,Arge10} and the
studies related to flare and filament eruptions \cite{Luna18}.   
In addition, maps  of the invisible-side (farside) of 
the Sun, computed from GONG Dopplergrams, identifying active regions before they rotate 
towards Earth are another important data product 
\cite{igh07,igh10}. An example of consequences of low-duty cycle in farside mapping is
illustrated in Figure~\ref{farside}. Here we display farside maps for three 
different duty cycles using 24-h GONG observations for June 21, 2014. Top panel shows 
the map from 100\,\% duty cycle as observed and other two panels display maps for 
lower duty cycles by artificially introducing gaps in the observations. It is clearly 
visible that reducing duty cycle reduces signal to noise and also increases random noise.
 It is also important
to note that the uninterrupted synoptic observations are crucial in advancing our knowledge
of the phenomena related to solar activity whose origin is still not fully understood
\cite{Elsworth15}.

\section{Summary}
\label{S-Summary}
Continuous observations of the Sun play an important role in the studies of long-term
as well as short-term solar variability. In this context, GONG comprising of six
ground-based observatories has been providing consistent and continuous observations 
spanning over more  than two solar cycles with a stable temporal coverage. 
 In this paper, we perform a detailed analysis of all the six sites and assess the 
 duty cycle that has been achieved by individual as well as combined sites. 
 
 The study reveals that the maximum observations have been recorded at two sites 
 located in the southern hemisphere and the lowest number of observations were 
 obtained at the UD site, predominantly owing to the yearly site closure during the 
 Indian monsoon. Further, there is significant overlap between observations from different 
sites, which is important for better calibration and for creating a seamless, long time 
series by merging observations from various sites. The mean and median duty cycle of the
network is 88\,\% and 92\,\%\,, respectively,  

 Our study further demonstrates that strong seasonal trends are present at all sites except 
 the Mauna Loa (ML) site, probably due to its geographical location.  However, these trends 
 disappear in the network when the observations from all sites are combined/merged. 
 Although the down time of each site varies from year to year, we do not  find  
 long-term change in climate at any of the network sites that may affect the duty cycle.

It should be noted that the ground-based networks are less expensive as compared to space 
missions and have the ease of repairing/upgrading any instrument though the image quality 
might be affected by the atmospheric seeing. The detailed analysis presented here 
demonstrates that a significantly high duty cycle values can be achieved from a 
well-designed network of ground-based  observatories and  can provide an important 
platform for long-term synoptic observations.

\ack

The authors would like to thank Jack Harvey, John Leibacher, and Charles Lindsey for 
critically reading the manuscript and their suggestions. K. Jain 
thanks Valent\'in Mart\'inez 
Pillet for many useful discussions. GONG has received support from hundreds of people at
the GONG headquarters at the National Solar Observatory and around the world,  and their 
dedication is pivotal for maintaining the GONG operations for such a long time. The authors 
would like to thank all those who are, or have been, associated with GONG from its planning 
to the present. This work utilizes data from the National Solar Observatory Integrated 
Synoptic Program,  which is operated by the Association of Universities for Research in 
Astronomy, under a cooperative agreement with the National Science Foundation and with 
additional financial support from the National Oceanic and Atmospheric Administration, 
the National Aeronautics and Space Administration, and the United States Air Force. The 
GONG network of instruments  is hosted by the Big Bear Solar Observatory, High Altitude 
Observatory, Learmonth Solar Observatory, Udaipur Solar Observatory, Instituto de 
Astrof\'{\i}sica de Canarias, and the Cerro Tololo Interamerican Observatory.

\References
\bibliographystyle{jphysicsB}

\harvarditem{{Arge} et~al.}{2010}{Arge10}
{Arge} C~N, {Henney} C~J, {Koller} J, {Compeau} C~R, {Young} S, {MacKenzie} D,
  {Fay} A \harvardand\ {Harvey} J~W  2010 {\em in} M~{Maksimovic},
  K~{Issautier}, N~{Meyer-Vernet}, M~{Moncuquet} \harvardand\ F~{Pantellini},
  eds, `Twelfth International Solar Wind Conference' Vol. 1216 of {\em American
  Institute of Physics Conference Series} p.~343.

\harvarditem{{Basu} \harvardand\ {Antia}}{2019}{Basu19}
{Basu} S \harvardand\ {Antia} H~M  2019 {\em \APJ} {\bf 883},~93.

\harvarditem{{Broomhall}}{2017}{Broomhall17}
{Broomhall} A~M  2017 {\em Sol Phys} {\bf 292},~67.

\harvarditem{{Chaplin} et~al.}{1996}{Chaplin96}
{Chaplin} W~J, {Elsworth} Y, {Howe} R, {Isaak} G~R, {McLeod} C~P, {Miller} B~A,
  {van der Raay} H~B, {Wheeler} S~J \harvardand\ {New} R  1996 {\em Sol Phys}
  {\bf 168},~1.

\harvarditem{{Chou} et~al.}{1995}{Chou95}
{Chou} D~Y, {Sun} M~T, {Huang} T~Y, {Lai} S~P, {Chi} P~J, {Ou} K~T, {Wang} C~C,
  {Lu} J~Y, {Chu} A~L, {Niu} C~S, {Mu} T~M, {Chen} K~R, {Chou} Y~P, {Jimenez}
  A, {Rabello-Soares} M~C, {Chao} H, {Ai} G, {Wang} G~P, {Zirin} H, {Marquette}
  W \harvardand\ {Nenow} J  1995 {\em Sol Phys} {\bf 160},~237.

\harvarditem{{Clark} et~al.}{2004}{Clark04}
{Clark} R, {Toner} C, {Hill} F, {Hanna} K, {Ladd} G, {Komm} R, {Howe} R,
  {Gonzalez-Hernandez} I \harvardand\ {Kholikov} S  2004 {\em in} D~{Danesy},
  ed., `SOHO 14 Helio- and Asteroseismology: Towards a Golden Future' Vol. 559
  of {\em ESA Special Publication} ESA Noordwijk p.~381.

\harvarditem{{Elsworth} et~al.}{2015}{Elsworth15}
{Elsworth} Y, {Broomhall} A~M, {Gosain} S, {Roth} M, {Jefferies} S~M
  \harvardand\ {Hill} F  2015 {\em Space Sci. Rev} {\bf 196},~137.

\harvarditem{{Finsterle} et~al.}{2004}{Finsterle04}
{Finsterle} W, {Jefferies} S~M, {Cacciani} A, {Rapex} P, {Giebink} C, {Knox} A
  \harvardand\ {Dimartino} V  2004 {\em Sol Phys} {\bf 220},~317.

\harvarditem{{Fischer} et~al.}{1986}{Fischer86}
{Fischer} G, {Hill} F, {Jones} W, {Leibacher} J, {McCurnin} W, {Stebbins} R
  \harvardand\ {Wagner} J  1986 {\em Sol Phys} {\bf 103},~33.

\harvarditem{{Fossat}}{1991}{Fossat91}
{Fossat} E  1991 {\em Sol Phys} {\bf 133},~1.

\harvarditem{{Fossat}}{2013}{Fossat13}
{Fossat} E  2013 {\em in} K~{Jain}, S.~C {Tripathy}, F~{Hill}, J.~W {Leibacher}
  \harvardand\ A.~A {Pevtsov}, eds, `Fifty Years of Seismology of the Sun and
  Stars' Vol. CS-478 of {\em Astron. Soc. Pacific Conf. Ser} San Francisco
  p.~73.

\harvarditem{{Gizon} et~al.}{2020}{Gizon20}
{Gizon} L, {Cameron} R~H, {Pourabdian} M, {Liang} Z~C, {Fournier} D, {Birch}
  A~C \harvardand\ {Hanson} C~S  2020 {\em Science} {\bf 368},~1469.

\harvarditem{{Gonz{\'a}lez Hern{\'a}ndez} et~al.}{2007}{igh07}
{Gonz{\'a}lez Hern{\'a}ndez} I, {Hill} F \harvardand\ {Lindsey} C  2007 {\em
  \APJ} {\bf 669},~1382.

\harvarditem{{Gonz{\'a}lez Hern{\'a}ndez} et~al.}{2010}{igh10}
{Gonz{\'a}lez Hern{\'a}ndez} I, {Hill} F, {Scherrer} P~H, {Lindsey} C
  \harvardand\ {Braun} D~C  2010 {\em Space Weather} {\bf 8},~06002.

\harvarditem{{Grec} et~al.}{1980}{Grec80}
{Grec} G, {Fossat} E \harvardand\ {Pomerantz} M  1980 {\em Nature} {\bf
  288},~541--544.

\harvarditem{{Hale} et~al.}{2016}{Hale16}
{Hale} S~J, {Howe} R, {Chaplin} W~J, {Davies} G~R \harvardand\ {Elsworth} Y~P
  2016 {\em Sol Phys} {\bf 291},~1--28.

\harvarditem{{Harvey} \harvardand\ {GONG Instrument Team}}{1995}{harvey95}
{Harvey} J \harvardand\ {GONG Instrument Team}  1995 {\em in} R.~K {Ulrich},
  E~{Rhodes}, Jr. \harvardand\ W~{D\"appen}, eds, `GONG 1994. Helio- and
  Astro-Seismology from the Earth and Space' Vol. CS-76 of {\em Astron. Soc.
  Pacific Conf. Ser} San Francisco p.~432.

\harvarditem{{Harvey} et~al.}{1998}{Harvey98}
{Harvey} J, {Tucker} R \harvardand\ {Britanik} L  1998 {\em in} S~{Korzennik},
  ed., `Structure and Dynamics of the Interior of the Sun and Sun-like Stars'
  Vol. 418 of {\em ESA Special Publication} Noordwijk p.~209.

\harvarditem{{Harvey}}{2013}{Harvey13}
{Harvey} J~W  2013 {\em in} K~{Jain}, S.~C {Tripathy}, F~{Hill}, J.~W
  {Leibacher} \harvardand\ A.~A {Pevtsov}, eds, `Fifty Years of Seismology of
  the Sun and Stars' Vol. CS-478 of {\em Astron. Soc. Pacific Conf. Ser} San
  Francisco p.~51.

\harvarditem{{Harvey} et~al.}{2011}{Harvey11}
{Harvey} J~W, {Bolding} J, {Clark} R, {Hauth} D, {Hill} F, {Kroll} R, {Luis} G,
  {Mills} N, {Purdy} T, {Henney} C, {Holland} D \harvardand\ {Winter} J  2011
  {\em in} `AAS/Solar Physics Division Abstracts \#42' Vol.~42 of {\em
  AAS/Solar Physics Division Meeting} p.~17.45.

\harvarditem{{Harvey} et~al.}{1996}{Harvey96}
{Harvey} J~W, {Hill} F, {Hubbard} R~P, {Kennedy} J~R, {Leibacher} J~W, {Pintar}
  J~A, {Gilman} P~A, {Noyes} R~W, {Title} A~M, {Toomre} J, {Ulrich} R~K,
  {Bhatnagar} A, {Kennewell} J~A, {Marquette} W, {Patron} J, {Saa} O
  \harvardand\ {Yasukawa} E  1996 {\em Science} {\bf 272},~1284--1286.

\harvarditem{{Hill}}{1988}{Hill88}
{Hill} F  1988 {\em \APJ} {\bf 333},~996.

\harvarditem{{Hill}}{2018}{Hill18}
{Hill} F  2018 {\em Space Weather} {\bf 16},~1488.

\harvarditem{{Hill} et~al.}{2008}{Hill08}
{Hill} F, {Bolding} J, {Clark} R, {Donaldson-Hanna} K, {Harvey} J~W, {Petrie}
  G~J~D, {Toner} C~G \harvardand\ {Wentzel} T~M  2008 {\em in} R~{Howe}, R.~W
  {Komm}, K.~S {Balasubramaniam} \harvardand\ G.~J.~D {Petrie}, eds,
  `Subsurface and Atmospheric Influences on Solar Activity' Vol. CS-383 of {\em
  Astron. Soc. Pacific Conf. Ser.} San Francisco p.~227.

\harvarditem{{Hill}, {Fischer}, {Forgach}, {Grier}, {Leibacher}, {Jones},
  {Jones}, {Kupke}, {Stebbins}, {Clay}, {Ingram}, {Libbrecht}, {Zirin},
  {Ulrichi}, {Websteri}, {Hieda}, {Labonte}, {Lu}, {Sousa}, {Garcia},
  {Yasukawa}, {Kennewell}, {Cole}, {Zhen}, {Su-Min}, {Bhatnagar}, {Ambastha},
  {Al-Khashlan}, {Abdul-Samad}, {Benkhaldoun}, {Kadiri}, {S{\'a}nchez},
  {Pall{\'e}}, {Duhalde}, {Solis}, {Sa{\'a}} \harvardand\
  {Gonz{\'a}lez}}{1994}{Hill94b}
{Hill} F, {Fischer} G, {Forgach} S, {Grier} J, {Leibacher} J~W, {Jones} H~P,
  {Jones} P~B, {Kupke} R, {Stebbins} R~T, {Clay} D~W, {Ingram} R~E~L,
  {Libbrecht} K~G, {Zirin} H, {Ulrichi} R~K, {Websteri} L, {Hieda} L~S,
  {Labonte} B~J, {Lu} W~M~T, {Sousa} E~M, {Garcia} C~J, {Yasukawa} E~A,
  {Kennewell} J~A, {Cole} D~G, {Zhen} H, {Su-Min} X, {Bhatnagar} A, {Ambastha}
  A, {Al-Khashlan} A~S, {Abdul-Samad} M~S, {Benkhaldoun} Z, {Kadiri} S,
  {S{\'a}nchez} F, {Pall{\'e}} P~L, {Duhalde} O, {Solis} H, {Sa{\'a}} O
  \harvardand\ {Gonz{\'a}lez} R  1994 {\em Sol Phys} {\bf 152},~351.

\harvarditem{{Hill}, {Fischer}, {Grier}, {Leibacher}, {Jones}, {Jones}, {Kupke}
  \harvardand\ {Stebbins}}{1994}{Hill94a}
{Hill} F, {Fischer} G, {Grier} J, {Leibacher} J~W, {Jones} H~B, {Jones} P~P,
  {Kupke} R \harvardand\ {Stebbins} R~T  1994 {\em Sol Phys} {\bf 152},~321.

\harvarditem{{Hill} \harvardand\ {Newkirk}}{1985}{Hill85}
{Hill} F \harvardand\ {Newkirk}, Jr. G  1985 {\em Sol Phys} {\bf 95},~201.

\harvarditem{{Howe} et~al.}{2020}{Howe20}
{Howe} R, {Chaplin} W~J, {Basu} S, {Ball} W~H, {Davies} G~R, {Elsworth} Y,
  {Hale} S~J, {Miglio} A, {Nielsen} M~B \harvardand\ {Viani} L~S  2020 {\em
  Monthly Notices Royal Astron Soc} {\bf 493},~L49.

\harvarditem{{Howe} et~al.}{2000}{Howe00}
{Howe} R, {Christensen-Dalsgaard} J, {Hill} F, {Komm} R~W, {Larsen} R~M,
  {Schou} J, {Thompson} M~J \harvardand\ {Toomre} J  2000 {\em Science} {\bf
  287},~2456.

\harvarditem{{Howe} et~al.}{2018}{Howe18b}
{Howe} R, {Hill} F, {Komm} R, {Chaplin} W~J, {Elsworth} Y, {Davies} G~R,
  {Schou} J \harvardand\ {Thompson} M~J  2018 {\em \APJL} {\bf 862},~L5.

\harvarditem{{Jackiewicz} et~al.}{2015}{Jackiewicz15}
{Jackiewicz} J, {Serebryanskiy} A \harvardand\ {Kholikov} S  2015 {\em \APJ}
  {\bf 805},~133.

\harvarditem{{Jain} et~al.}{2009}{Jain09}
{Jain} K, {Tripathy} S~C \harvardand\ {Hill} F  2009 {\em \APJ} {\bf
  695},~1567.

\harvarditem{{Jain} et~al.}{2011}{Jain11}
{Jain} K, {Tripathy} S~C \harvardand\ {Hill} F  2011 {\em \APJ} {\bf 739},~6.

\harvarditem{{Jefferies} et~al.}{2019}{Jefferies19}
{Jefferies} S~M, {Fleck} B, {Murphy} N \harvardand\ {Berrilli} F  2019 {\em
  \APJL} {\bf 884},~L8.

\harvarditem{{Komm} et~al.}{2018}{Komm18}
{Komm} R, {Howe} R \harvardand\ {Hill} F  2018 {\em Sol Phys} {\bf 293},~145.

\harvarditem{{Leibacher} \harvardand\ {GONG Project Team}}{1995}{Leibacher95}
{Leibacher} J \harvardand\ {GONG Project Team}  1995 {\em in} R.~K {Ulrich},
  E.~J {Rhodes}, Jr. \harvardand\ W~{D\"appen}, eds, `GONG 1994. Helio- and
  Astro-Seismology from the Earth and Space' Vol. CS-76 of {\em Astron. Soc.
  Pacific Conf. Ser.} San Francisco p.~381.

\harvarditem{{Leibacher} \harvardand\ {Stein}}{1971}{Leibacher71}
{Leibacher} J~W \harvardand\ {Stein} R~F  1971 {\em Astrophys Lett} {\bf
  7},~191.

\harvarditem{{Leighton}}{1960}{leighton60}
{Leighton} R~B  1960 {\em in} R.~N {Thomas}, ed., `Aerodynamic Phenomena in
  Stellar Atmospheres' Vol.~12 of {\em IAU Symp.} p.~321.

\harvarditem{{Leighton} et~al.}{1962}{Leighton62}
{Leighton} R~B, {Noyes} R~W \harvardand\ {Simon} G~W  1962 {\em \APJ} {\bf
  135},~474.

\harvarditem{{Luna} et~al.}{2018}{Luna18}
{Luna} M, {Karpen} J, {Ballester} J~L, {Muglach} K, {Terradas} J, {Kucera} T
  \harvardand\ {Gilbert} H  2018 {\em \APJS} {\bf 236},~35.

\harvarditem{{Pall{\'e}} et~al.}{2015}{Palle15}
{Pall{\'e}} P~L, {Appourchaux} T, {Christensen-Dalsgaard} J \harvardand\
  {Garcx{\'\i}a} I~A  2015 {\em in} V.~C.~H {Tong} \harvardand\ R.~A {García},
  eds, `Extraterrestrial Seismology' Cambridge University Press p.~25.

\harvarditem{{Petrie} et~al.}{2008}{Petrie08}
{Petrie} G~J~D, {Bolding} J, {Clark} R, {Donaldson-Hanna} K, {Harvey} J~W,
  {Hill} F, {Toner} C \harvardand\ {Wentzel} T~M  2008 {\em in} R~{Howe}, R.~W
  {Komm}, K.~S {Balasubramaniam} \harvardand\ G.~J.~D {Petrie}, eds,
  `Subsurface and Atmospheric Influences on Solar Activity' Vol. CS-383 of {\em
  Astron. Soc. Pacific Conf. Ser.} San Francisco p.~181.

\harvarditem{{Pintar} \harvardand\ {Toussaint}}{1998}{Pintar98}
{Pintar} J \harvardand\ {Toussaint} R  1998 {\em in} S~{Korzennik}, ed.,
  `Structure and Dynamics of the Interior of the Sun and Sun-like Stars' Vol.
  418 of {\em ESA Special Publication} p.~295.

\harvarditem{{Simoniello} et~al.}{2013}{Simoniello13a}
{Simoniello} R, {Jain} K, {Tripathy} S~C, {Turck-Chi{\`e}ze} S, {Baldner} C,
  {Finsterle} W, {Hill} F \harvardand\ {Roth} M  2013 {\em \APJ} {\bf
  765},~100.

\harvarditem{{Toner}}{2001}{Toner01}
{Toner} C~G  2001 {\em in} A~{Wilson} \harvardand\ P.~L {Pall{\'e}}, eds, `SOHO
  10/GONG 2000 Workshop: Helio- and Asteroseismology at the Dawn of the
  Millennium' Vol. 464 of {\em ESA Special Publication} Noordwijk p.~355.

\harvarditem{{Toner} et~al.}{2003}{Toner03}
{Toner} C~G, {Haber} D, {Corbard} T, {Bogart} R \harvardand\ {Hindman} B  2003
  {\em in} H~{Sawaya-Lacoste}, ed., `GONG+ 2002. Local and Global
  Helioseismology: the Present and Future' Vol. 517 of {\em ESA Special
  Publication} Noordwijk p.~405.

\harvarditem{{Toussaint} et~al.}{1995}{Toussaint95}
{Toussaint} R, {Harvey} J \harvardand\ {Hubbard} R  1995 {\em in} R.~K
  {Ulrich}, E.~J {Rhodes}, Jr. \harvardand\ W~{D\"appen}, eds, `GONG 1994.
  Helio- and Astro-Seismology from the Earth and Space' Vol. CS-76 of {\em
  Astron. Soc. Pacific Conf. Ser.} p.~532.

\harvarditem{{Tripathy} et~al.}{2013}{Tripathy13}
{Tripathy} S~C, {Jain} K \harvardand\ {Hill} F  2013 {\em Sol Phys} {\bf
  282},~1.

\harvarditem{{Tripathy} et~al.}{2015}{Tripathy15}
{Tripathy} S~C, {Jain} K \harvardand\ {Hill} F  2015 {\em \APJ} {\bf 812},~20.

\harvarditem{{Ulrich}}{1970}{Ulrich70}
{Ulrich} R~K  1970 {\em \APJ} {\bf 162},~993.

\harvarditem{{Woodard} \harvardand\ {Noyes}}{1985}{Woodard85}
{Woodard} M~F \harvardand\ {Noyes} R~W  1985 {\em Nature} {\bf 318},~449.

\endrefs
\end{document}